\documentclass[manuscript]{emulateapj}

\def\msun{M$_{\odot}$}

\def\den{g cm$^{-3}$}

\begin{document}

\title{Magnetohydrodynamic Simulations of Disk Galaxy Formation: \\ the Magnetization of The Cold and Warm Medium}

\author{Peng Wang and Tom Abel}
\affil{1. Kavli Institute for Particle Astrophysics and Cosmology,
  \\ Stanford Linear Accelerator Center and Stanford Physics Department, Menlo Park, CA 94025\\
  2. Kavli Institute for Theoretical Physics, University of California, Santa Barbara, CA 93106} 
\email{pengwang,
  tabel@stanford.edu}

\begin{abstract}
Using magnetohydrodynamic (MHD) adaptive mesh refinement simulations,
we study the formation and early evolution of
disk galaxies with a magnetized interstellar medium. For a $10^{10}$
\msun \ halo with initial NFW dark matter and gas profiles, we impose
a uniform $10^{-9}$ G magnetic field and follow its collapse, disk
formation and evolution up to 1 Gyr. Comparing to a purely
hydrodynamic simulation with the same initial condition, we find that
a protogalactic field of this strength does not significantly
influence the global disk properties. At the same time, the initial
magnetic fields are quickly amplified by the differentially rotating
turbulent disk. After the initial rapid amplification lasting
$\sim500$ Myr, subsequent field amplification appears
self-regulated. As a result, highly magnetized material begin to form
above and below the disk. Interestingly, the field strengths in the
self-regulated regime agrees well with the observed fields in the
Milky Way galaxy both in the warm and the cold HI phase and do not
change appreciably with time. Most of the cold phase shows a
dispersion of order ten in the magnetic field strength. The global
azimuthal magnetic fields reverse at different radii and the amplitude
declines as a function of radius of the disk. By comparing the
estimated star formation rate (SFR) in hydrodynamic and MHD
simulations, we find that after the magnetic field strength saturates,
magnetic forces provide further support in the cold gas and lead to a
decline of the SFR. 
\end{abstract}

\maketitle

\section{Introduction}

Rapid developments in observational cosmology recently established a standard model of cosmology, the $\Lambda$CDM model \citep{spergel07}. In principle, this makes the problem of galaxy formation essentially an initial value problem. Indeed, Gpc-scale N-body simulations of $\Lambda$CDM cosmology are revealing the details of large-scale structure formation and clustering, matching observations in many aspects \citep{springel05}. In this framework, it should be possible to understand the formation and evolution processes of individual galaxies. For this purpose, it is crucial to understand the structure of interstellar medium (ISM) and the dominant physics controlling star formation.

Analytical and semi-analytical studies of disk galaxy formation are
useful in understanding the global properties \citep{mo98,
  efstathiou00, dutton07, silk07, stringer07}. Because of the complex
nature of galaxy formation and the wide range of physical processes
involved, numerical simulations also prove particularly
useful. Hydrodynamical simulations of galaxy formation in both
cosmological and isolated setups have taught us a lot about the
detailed structure of the galactic ISM and star formation (see
e.g. Kravtsov 2003, Li et al. 2005, Springel \& Hernquist 2005, Okamoto et al. 2005, Tasker
\& Bryan 2006, Tassis et al. 2006, Governato et al. 2007, Wada \&
Norman 2007, Kaufmann et al. 2007, Robertson \& Kravtsov 2007 for
recent studies). However, there are still many physical processes we need to include to make the simulations realistic. 

In this work, we focus on magnetic fields which have not yet been included in previous galaxy formation simulations.  Magnetic fields may significantly
influence the structure and evolution of ISM and has been studied
extensively previously by local magnetohydrodynamic (MHD) simulations
of the ISM \citep{maclow05, balsara04, avillez05, piontek07,
  hennebelle06} and by observations (e.g. Beck 2007, Crutcher
1999). On a larger scale, the effect of magnetic fields is less clear
because cosmological MHD simulations are still at an early
stage. \citet{miniati01} performed cosmological hydrodynamic
simulations for a passive magnetic field. Cosmological MHD simulations
have been done in SPH (e.g. Dolag 1999) in the context of galaxy
cluster formation and Eulerian-code simulations are also beginning to be
explored (e.g. Li et al. 2006). None of those calculations have as yet
resolved the internal structures of disk galaxies.

Besides the potential importance in the dynamics of galaxy formation,
galactic magnetic fields are a key ingredient in many other
astrophysical problems. For example, they play a dominant role in the
origin and propagation of cosmic rays \citep{fermi49, strong07}, which
may also be dynamically important in galaxy formation \citep{fermi54,
  ensslin07}. Furthermore, magnetic fields may be important in the
dynamics of supernova remnants \citep{balsara01a}, in
regulating the internal turbulence of giant molecular clouds by
driving outflows \citep{zhiyun06, banerjee07} and in the collapse of
individual molecular cloud cores \citep{mouschovias87, balsara01b,
  tilley07, mellon07}.

While the importance on the ISM is generally accepted, the origin of
the galactic magnetic field is still an unsolved problem (see
e.g. Kulsrud \& Zweibel 2007 for a recent review). The traditional
theory of a galactic dynamo, the alpha-Omega dynamo, is based on the
mean field dynamo theory introduced by \citet{steenbeck66} and later
developed by \citet{parker71} and \citet{vainshtein71}.
\citet{ferriere92} has extended this galactic dynamo theory to include
the expansion of supernova bubbles. Turbulent amplification in
protogalaxies have also been studied \citep{pudritz89,
  kulsrud97}. Those theories are built on
kinematic dynamo. A different paradigm for disk dynamo comes with the
discovery of the magneto-rotational instability (MRI)
\citep{chandra61, balbus91}. In MRI, field amplification is not only a
consequence of the turbulent velocity field, it also produces the
turbulent velocity field itself. It is still unclear what is the
dominant mechanism that amplifies the galactic magnetic
field. In contrast to the case of dynamos in planets,
because of the enormous inductance of the ISM \citep{fermi49}, there
is no decaying problem for the galactic field. So the essential part
of the problem here is indeed only the initial amplification.

In the following, we first describe the numerical algorithms and
initial conditions. Then in section~3 we present the
results. Sections~4 and 5 are devoted to discussion and conclusions, respectively.

\section{Numerical Methods and Models}

\subsection{MHD Algorithm}

We have developed a 3D adaptive mesh refinement (AMR) MHD code based
on the cosmological AMR hydrodynamics code Enzo \citep{bryan97,
  o'shea04}. We use the Dedner formulation of MHD equations
\citep{dedner}. This conservative formulation of the MHD equations
uses hyperbolic divergence cleaning. It has also been used in several
other recent AMR MHD codes \citep{matsumoto, anderson}. The
conservative system is discretized by method of lines. The Riemann
solver and reconstruction schemes have been tested extensively in
\citet{wang07} for relativistic systems. Since our Riemann solvers and
reconstruction schemes are designed to work for any conservative
systems, when applying it to the Dedner MHD equations, we only 
redefine the conservative variables. Using these
high-resolution-shock-capturing (HRSC) schemes designed for
conservative systems ensures the conservation of mass, momentum and
energy, as well as obtaining correct shock positions and velocities
\citep{leveque}. Interested readers can find more detailed
descriptions of our numerical algorithms in \citet{wang07} and
references therein. In this work, reconstruction is done using the
piecewise linear method (PLM) \citep{van leer79}. Fluxes at cell
interfaces are calculated using the local Lax-Friedrichs Riemann
solver (LLF, Kurganov \& Tadmor 2000), which is free of the carbuncle
artifacts present in some other contact-capturing Riemann solvers
\citep{quirk}. Time integration is performed using the total variation
diminishing (TVD) second order Runge-Kutta (RK) scheme
\citep{shu88}. The code inherits all of Enzo's parallel AMR capability
as well as self-gravity, chemistry, cooling and radiative transfer
routines. We have tested the code using various standard hydrodynamics
and MHD test problems including the 1D Brio-Wu problem, 2D MHD rotor,
2D Hydro and MHD Rayleigh-Taylor problem, 3D Sedov-Taylor problem, 3D
Larson-Penston isothermal collapse problem, etc.

\subsection{Cooling}

A cooling function is used to calculate the radiative energy losses down to temperature $300$~K. We use the cooling function fitted by \citet{sarazin87} for $T>10^4 \ {\rm K}$ and by \citet{rosen} for $300 \ {\rm K}<T<10^4$ K. 

In global simulations of galactic disks, current computational power
does not allow to resolve the Jeans length down to molecular core
scales where star formation actually occurs. This may gives rise to
artificial fragmentation if one underresolves the Jeans length
\citep{truelove97}. This potentially worry is avoided using a density
dependent temperature floor ensuring that a cell always resolves half
of the local Jeans length (e.g. Wada \& Norman 2007):
\begin{equation}
T_{min}=284 {\rho\over10^{-22} \ {\rm g cm^{-3}}}\left({\Delta x\over 20 \ {\rm pc}}\right)^2 \ {\rm K} \ . 
\end{equation}
Another common approach would be to create a collisionless star particle if a cell fails to resolve Jeans length (e.g. Tasker \& Bryan 2006).

\begin{figure*}[!]
 % \plotone{f4.eps}
  \includegraphics[height=.65\textheight]{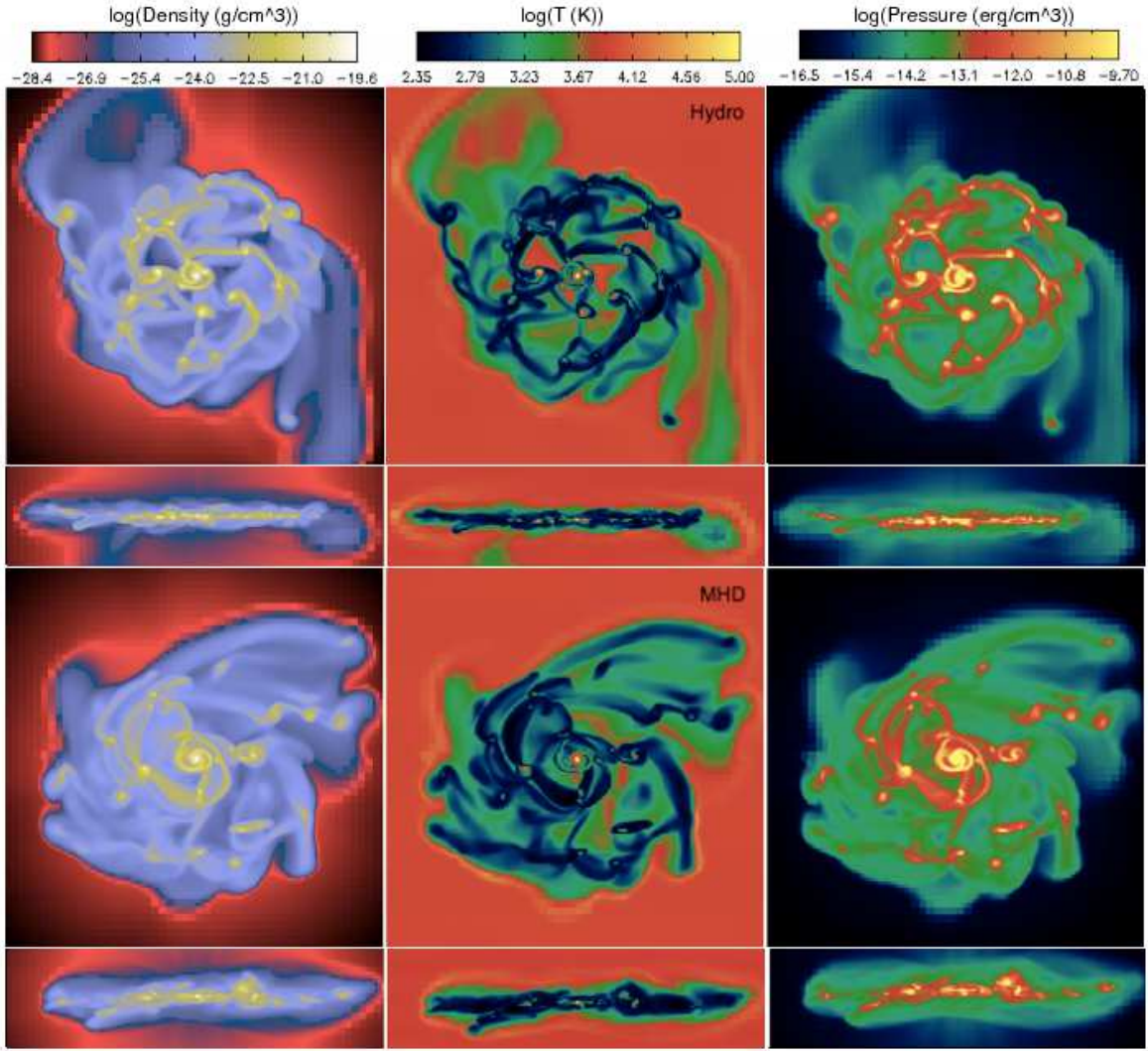}
    \caption{Face-on and edge-on density-weighted projections of
      density (left), temperature (middle) and thermal pressure
      (right) at $t=1.088$ Gyr for the Hydro (top two rows) and MHD (bottom two rows) runs. The physical length of the plot is $11$ kpc.} \label{proj}
\end{figure*}

\begin{figure*}[!]
 % \plotone{f4.eps}
  \includegraphics[height=.4\textheight]{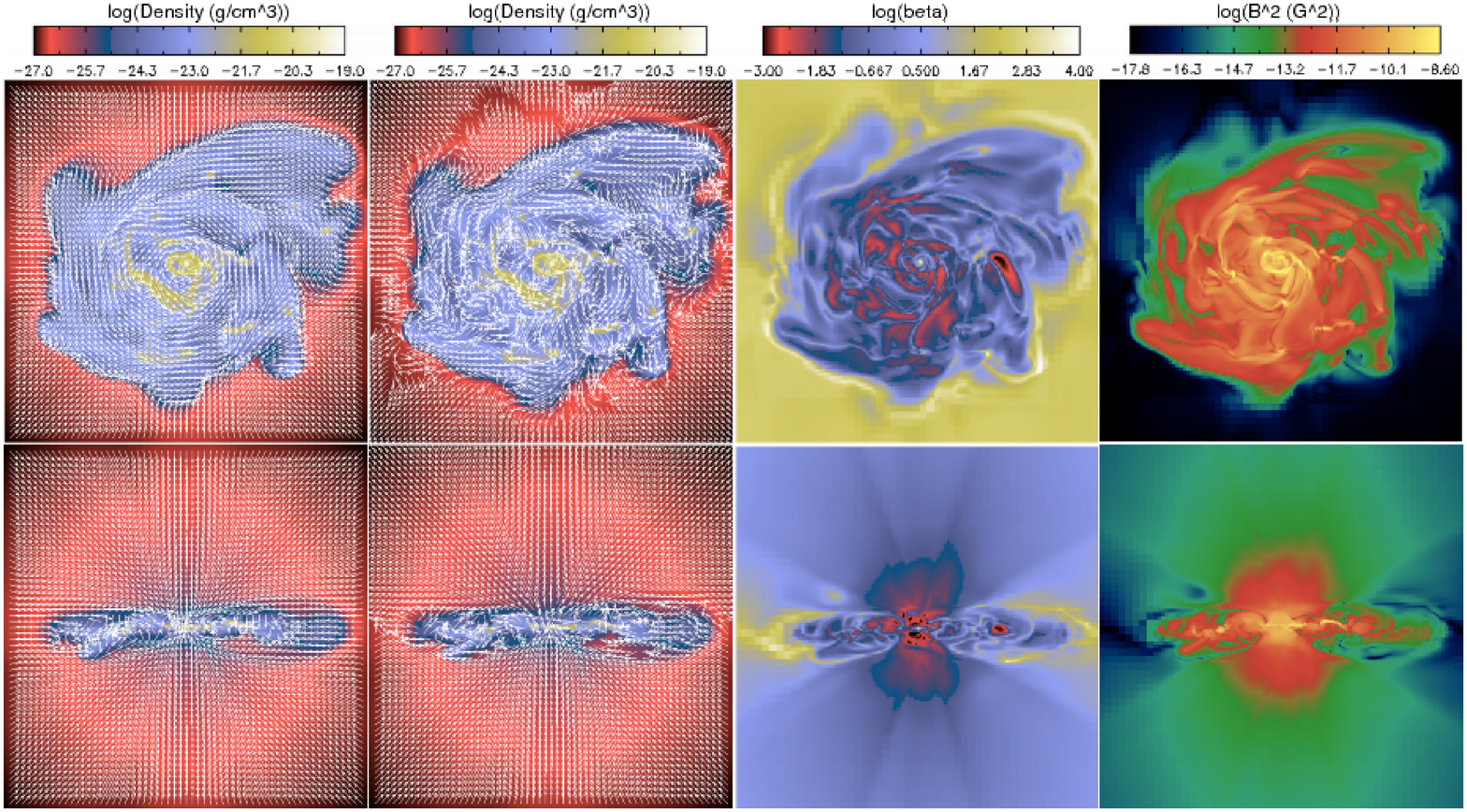}
    \caption{Velocity and Magnetic field plots at $t=1.088$ Gyr. From left to right: 1. velocity field vectors overplotted on a density slice; 2. magnetic
      field vectors overplotted on a density slice; 3. slice of
      plasma beta; 4. slice of $B^2$. All face-on slices are
      through $z=0.5$ and edge-on slices are through $y=0.5$. The
      physical length of the plot corresponds to $11$ kpc.} \label{bfield}
\end{figure*}

\subsection{Initial Conditions}

From pure N-body simulations it has been found that cold dark matter halos have a universal density profile well fitted by the NFW form \citep{nfw}. Here we consider a small halo formed at redshift $2$, with a mass of $M_{vir}=10^{10}$ \msun and correspondingly, a virial radius $R_{vir}=21.5$ kpc. The spin parameter is assumed to be $\lambda=0.05$ and concentration is $c_{vir}=10$. Such a small halo allows for higher spatial resolution. By using an isolated model, we are implicitly assuming that the effect of major and minor merger on the growth of galaxy disk is subdominant, which may be a reasonable assumption if the halo is smaller than the Milky Way halo \citep{guo07}. However, our model galaxy is still larger than the small galaxies with rotation velocities $<30$ km/s in which star formation may be strongly suppressed by the intergalactic UV background \citep{thoul96}. 
The gravitational interaction between dark matter and baryons is a secondary effect. Consequently in this work we model the dark matter halo as a static external potential. 

The initial gas density profile follows also the NFW profile with its
amplitude determined from the assumed baryon mass fraction
$f_b=0.1$. The initial gas temperature is set by solving for local
hydrostatic equilibrium. The rotation velocity is set to be the same
as that of the dark matter following \citet{springel99}. In addition,
we also add to the gas random velocities of the same order as the dark
matter virial velocity to model crudely the turbulent motions in
hierarchical formation of protogalaxies. We put the NFW halo at the
center of the simulation box with virial radius $0.1$ (21~kpc) of
the simulation box length. This makes sure that the gas evolution
inside halos will be minimally affected by numerical boundary
effects. The gas outside the halo virial radius is set to be the
cosmic mean value at redshift $2$ with a temperature of $T=10^4$
K. The topgrid resolution is $64^3$, and there are three static nested
grids refined by a factor of two around the halo. Then we let the code
dynamically refine to higher levels according to the Jeans criterion
with Jeans number 4. After the disk forms, almost all the disk is
refined to this highest level because it is strongly
self-gravitating. The simulations were run with a maximum refinement
level of six and seven, corresponding to resolutions of $52$ and  $26$
pc, respectively. We did not find noticeable differences in the
results. So in the following, only the results of the $26$ pc
resolution run are discussed.

The initial magnitude and topology of the magnetic field is
uncertain. Cosmological simulations for the generation of magnetic
seed field will be required to address this important question. A tiny
seed magnetic field of the order $10^{-21}$ G is always guaranteed by
the Biermann battery effect \citep{biermann50}. However, there are
both observational and theoretical arguments suggesting larger
pregalactic field strengths. For example, Faraday rotation has been
detected in high redshift damped Lyman alpha system
\citep{oren95}. Also, the abundance of beryllium and boron in old
Galactic halo stars is directly proportional to their iron
abundance. Big bang nucleosynthesis does not produce beryllium nor
boron. Hence, they may dominately  be produced by spallation of low
energy (tens of MeV) carbon and oxygen cosmic rays, suggesting that
magnetic fields and cosmic rays are already present at early times
\citep{zweibel03}. On a more theoretical ground, in hierarchical
structure formation, any halo formed at late times must have had
progenitors that hosted prior generations of stars. So any magnetic
field produced by those stars, their supernovae, or their pulsar remnants would be amplified by the turbulent ISM in the galaxies containing them. Turbulence driven by mergers may also amplify magnetic fields \citep{pudritz89, kulsrud92}. Furthermore, \citet{rees06} argued that the magnetic fields from supernova ejecta and extended radio lobes may build a field in excess of $10^{-9}$ G seed field for galactic disk forming in the late Universe. In this work, we adopt the simple assumptions that the initial magnetic field is weak, uniform and directed along the rotation axis. We will discuss the results of two simulations, with $B=0$ (Hydro run) and $B=10^{-9}$ G (MHD run) throughout. Initially, the magnetic field is dynamically unimportant, it will be twisted and amplified by the turbulent velocity field. 

This first study of MHD models of disk galaxy formation only include a
minimal set of physical processes relevant for the rich interplay
between magnetic fields, the dynamics of disk formation and a
multiphase ISM. We do not yet include supernova feedback, radiative
heating, cosmic rays, molecular cloud chemistry, ambipolar diffusion
nor other physical processes that may be relevant. With those
limitations in mind, our model should be viewed as a numerical
experiment rather than a realistic simulation of current day disk
galaxies. Some of those neglected feedback processes, e.g., supernova
feedback, may significantly change the conclusions drawing from the
current calculations.

\begin{figure*}[!]
 % \plotone{f4.eps}
  \includegraphics[height=.5\textheight]{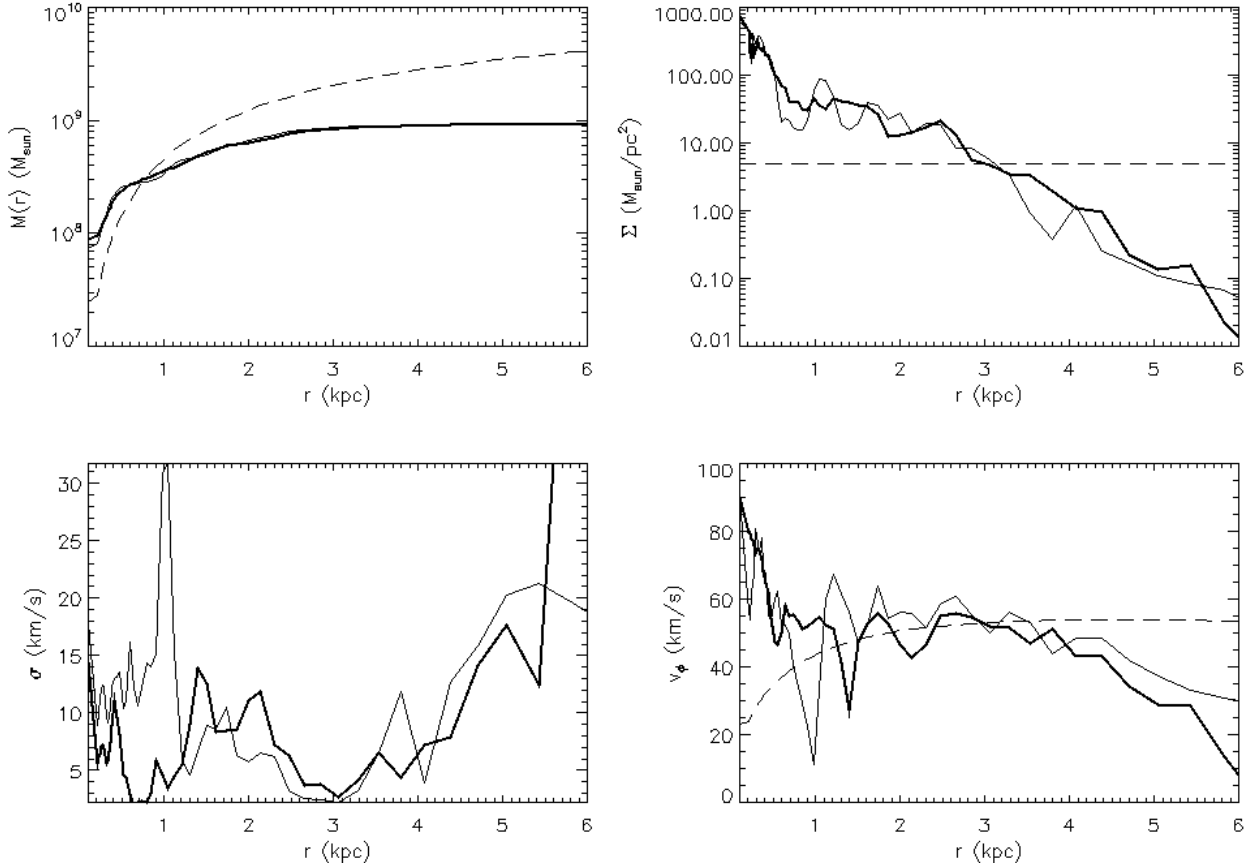}
    \caption{Disk profiles for the Hydro run (thin lines) and the MHD run (thick lines) at $t=1.088$. Top left gives the mass profile, top right the surface density profile, bottom left the velocity dispersion, bottom right the rotational velocity profile. In the mass profile plot, the thin dashed line is the mass profile of the static NFW halo. In the surface density profile plot, the dashed line is the threshold gas surface density for star formation $\approx5$ \msun pc$^{-2}$ as observed by \citet{martin01}. In the rotational velocity plot, the thin dashed line is the rotation velocity corresponding to the static NFW halo $v_{\phi, NFW}(r)=[GM_{NFW}(r)/r]^{1/2}$.} \label{disk}
\end{figure*}

\begin{figure}
 % \plotone{f4.eps}
  \includegraphics[height=.25\textheight]{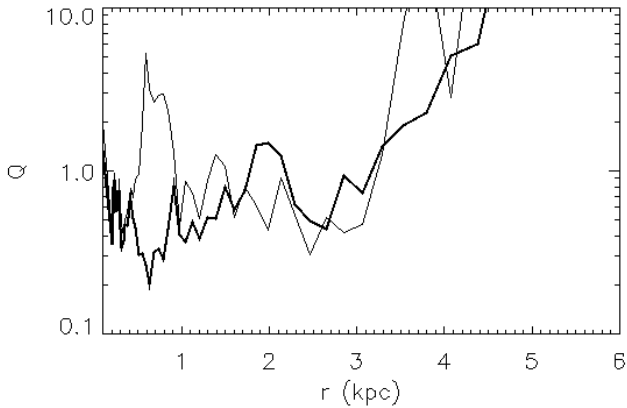}
    \caption{Toomre Q profiles for Hydro run (thin line) and MHD run (thick line) at $t=1.088$.} \label{q}
\end{figure}

\section{Results}

Fig.~\ref{proj} shows the face-on and edge-on projections of the density, temperature and the thermal pressure. Fig. \ref{bfield} shows face-on and edge on slices of velocity field and magnetic field lines overplotted on density, plasma beta $\beta=p/(B^2/(8\pi))$ and $B^2$. Both figures correspond to $t=1.088$~Gyr. In some of the plots below, we will consider only the gas in the disk defined by a density threshold criterion $\rho>1\times10^{-26}$ \den.

%\begin{figure*}[!]
 % \plotone{f4.eps}
 % \includegraphics[height=.5\textheight]{disk-b9.eps}
    %\caption{Disk profiles for MHD run at $t=272$ Myr (thin solid line), 544 Myr (dotted line), 680 Myr (dashed line), 
    %816 Myr (dash-dotted line) and 1.088 Gyr (solid line). Top left is mass-weighted averaged radial velocity, top right is mass-weighted averaged rotation velocity, bottom left is mass profile and bottom right is surface density profile. In the mass profile plot, the thin dashed line is the mass profile of the static NFW halo. In the rotational velocity plot, the thin dashed line is the rotation velocity corresponding to the static NFW halo $v_{\phi, NFW}(r)=(GM_{NFW}(r)/r)^{1/2}$.} \label{disk-b9}
%\end{figure*}

\subsection{Disk Structures} 

As can be seen in Fig.~\ref{proj}, the global disk structures are
qualitatively similar in both the Hydro and MHD cases. The density
projections show filamentary structures and high density ``blobs''
located along those filaments. Were we to Include molecular hydrogen chemistry
and cooling, these blobs would presumably resemble giant molecular
cloud complexes. This is similar to the density distributions of HI
gas and GMCs in observations of nearby galaxies such as M33
\citep{engargiola03}, which also show good correspondence between the
filaments and the locations of the GMCs. The temperature projections
show that those filaments are all cold, with $T\approx300$~K. The
lower density disk material has a temperature of $T\approx 10^4$ K. In
the MHD run, Fig.~\ref{bfield} shows that much of the cold gas is already
strongly magnetized.

In a self-gravitating disk, the vortex mode is the fastest unstable
mode with growth rates $2-3$ times larger than the spiral density wave mode \citep{mamatsashvili07}. In an isolated non-rotating fluid, the perturbation mode is divided into vortical and divergent types according to the Helmholz decomposition. And it is mainly the energy in divergent flows that will be dissipated in shocks. In a differentially rotating disk, the vortex mode is a combination of vortical and divergent motions. Thus the morphology and statistical properties of large scale shocks which form the filaments are different in those two cases. However, there seems to be no quantitative comparative studies on this problem that we are aware of. 

To see how self-gravity affects the growth of vortex modes, we have
carried out experiments with only the external NFW potential,
neglecting the self-gravity of the gas. Analytical calculations in the thin sheet approximation predict that the growth rate of vortex mode is orders of magnitude smaller without self-gravity \citep{mamatsashvili07}. Indeed in this case we saw that the disk shows
much less filamentary structures and instead develops tightly wound
spiral density wave patterns. Thus the vortical turbulent motion in
the fiducial cases is clearly driven by self-gravity.  Another
interesting implication of this experiment is that many local galaxies
do show tightly wound spiral patterns. Hence our experiment suggests
that the gravitational force in the gas disks of those galaxies is
dominated by the stars. This implies that for disk galaxies at high
redshift that are still self-gravitating, the vortex modes may be the
dominant structures regulating molecular cloud and star formation
within them.  In the following discussion we will call the turbulent
velocity field seen in our simulations gravity-driven turbulence
\citep{gammie01, wada02}.

As also can be seen in Fig.~\ref{proj}, the MHD run has a somewhat thicker disk than the Hydro run. This shows at $t=1.088$ Gry, magnetic pressure is already large enough to provide additional support to the vertical force balance, which can also be seen from Fig.~\ref{bfield}.

The pressure projections, show that the cold filaments have higher pressure than the warm gas and the cores of the filaments have even higher pressure. The overpressure in the cores is expected as they are self-gravitating \citep{larson81}. The higher pressure in non-selfgravitating cold gas is consistent with previous simulations of isolated disk galaxies \citep{tasker06}. This result is different from the idea that cold gas is formed by isobaric thermal instability in the galactic disk \citep{field65}. One important difference with the original thermal instability analysis is that instead of cosmic ray heating, the heating sources in our simulation are $pdV$ work, numerical viscous heating and numerical resistivity in the MHD run, all controlled by the gravity-driven turbulence velocity field in the disk. Thus in our simulations show that gravity-driven turbulent heating and radiative cooling will create a non-isobaric two phase medium. In the old isobaric theory, there is no prediction for the morphology and fraction of the cold phase. But in the turbulent heating scenario, one would expect the cold gas to have filamentary structures, which are naturally produced by a supersonic turbulent velocity field. 

Fig.~\ref{disk} shows the azimuthally averaged radial profiles of
mass, surface density, velocity dispersion and rotational velocity for
both simulations at $t=1.088$ Gyr. The Hydro and MHD calculations have
similar density profiles and remarkably similar mass profiles. This is
because magnetic field amplification happens after the disk is formed
so the initial weak magnetic field will not affect significantly the
dynamics of disk formation. In both cases, there are no sharp
boundaries at the disk edge in surface density profile. From the mass
profile one sees the disk radius to be $\sim3$ kpc for both simulations. We overplotted the threshold gas surface density for star formation $\approx5$ \msun pc$^{-2}$ derived from observations \citep{martin01}. It is interesting to see that it is just near the disk radius. In the mass profile plot, the dark matter halo mass profile is also shown. It shows that the dark matter mass dominates baryon at $r>1$ kpc. This explains why the rotational velocity profile is roughly flat at $r>1$ kpc with $v_\phi\approx50$ km/s, appropriate for the chosen NFW mass profile. In the rotation velocity plot, there is a big trough at $\sim 1$ kpc in both the Hydro and MHD curves. This is because of the fact that there are some big clumps at this radius whose spin provides a large cancellation to the mass-averaged rotation velocity. This also gives the trough in Toomre Q parameter discussed below.

The bottom left panel of Fig.~\ref{disk} shows the velocity dispersions calculated by $\sigma=\sqrt{<v^2>}-<v_{\phi}>$ where $<\cdot>$ represents mass-weighted azimuthal average. The velocity dispersion has been argued to remain roughly constant in self-regulated regions of disks \citep{silk97}, but this is still controversial \citep{ferguson98}. We can see that across the disk the velocity dispersions have about an order of magnitude fluctuations centered around $5-10$ km/s. So taking it to be constant is only a rough approximation. However, \citet{silk97} argued that the hot phase is the key in creating the self-regulated stage and determining the global velocity dispersions. Since hot phase is completely missing in our simulations, the situation might be different once supernova feedback is included. 

Fig.~\ref{q} shows the Toomre Q parameter $Q=\Omega \sigma/(\pi G\Sigma)$ \citep{toomre}, where $\Omega$ is the angular velocity, $\sigma$ is the velocity dispersion, $\Sigma$ is the surface density. In the original Toomre's formula, $\Omega$ is replaced by the epicyclic frequency $\kappa=(rd\Omega^2/dr+4\Omega^2)^{1/2}$. We used $\Omega$ because our disk is highly turbulent, different averaging methods for calculating $\Omega$ would give different $\kappa$. In practice $\kappa\sim\Omega$ holds to within a factor of two. As can be seen from Fig.~\ref{q}, throughout the disk Q fluctuates around $1$, implying that the disk is Toomre unstable. Such a roughly constant $Q$ suggests that the gravity-driven turbulence in the disk is in a quasi-stationary state at this time \citep{gammie01, wada02}. This is also supported by the quasi-stationary density PDF (see section \ref{ism}). Beyond the disk radius, $Q$ rises sharply to $>1$. 

%To see how the disk surface density profile evolves with time, we plotted the time evolution of surface density profile for the Hydro run in Fig. We can see that the surface density profile is building from inside-out and remains qusi-stationary after formation. The MHD run shows the same behavior.

\begin{figure}
 % \plotone{f4.eps}
  \includegraphics[height=.4\textheight]{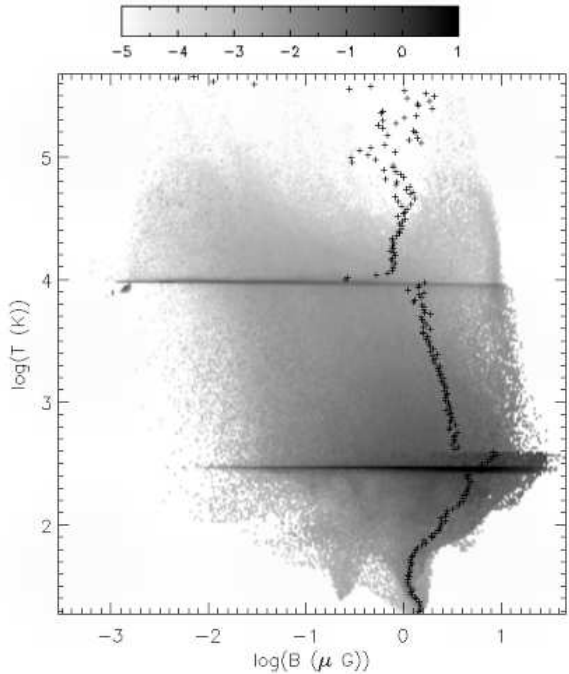}
    \caption{Mass-weighted joint temperature-B PDF at $t=1.088$ Gyr. The total mass in the disk is $\sim10^9$ \msun, as can be seen from the mass profile plot of Fig.~\ref{disk}. The binsize is 0.02 in log space. The plus signs are the averaged magnetic field strength in every temperature bins.} \label{tb}
\end{figure}

\begin{figure}
 % \plotone{f4.eps}
  \includegraphics[height=.35\textheight]{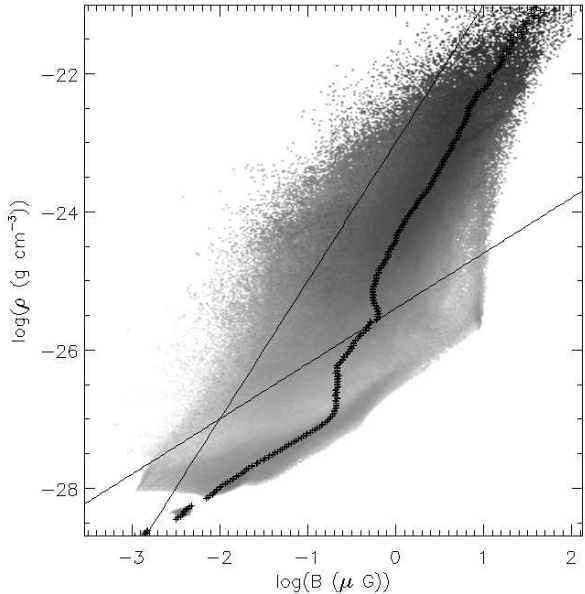}
    \caption{Mass-weighted joint density-B PDF at $t=1.088$ Gyr. The solid straight lines are $\rho\propto B^{2}$ and $\rho\propto B^{0.8}$. The color scale and binsize are the same as that of Fig. \ref{tb}. The plus signs are the averaged magnetic field strength in every density bins.} \label{rhob}
\end{figure}

\begin{figure}
 %\plotone{f4.eps}
  \includegraphics[height=.3\textheight]{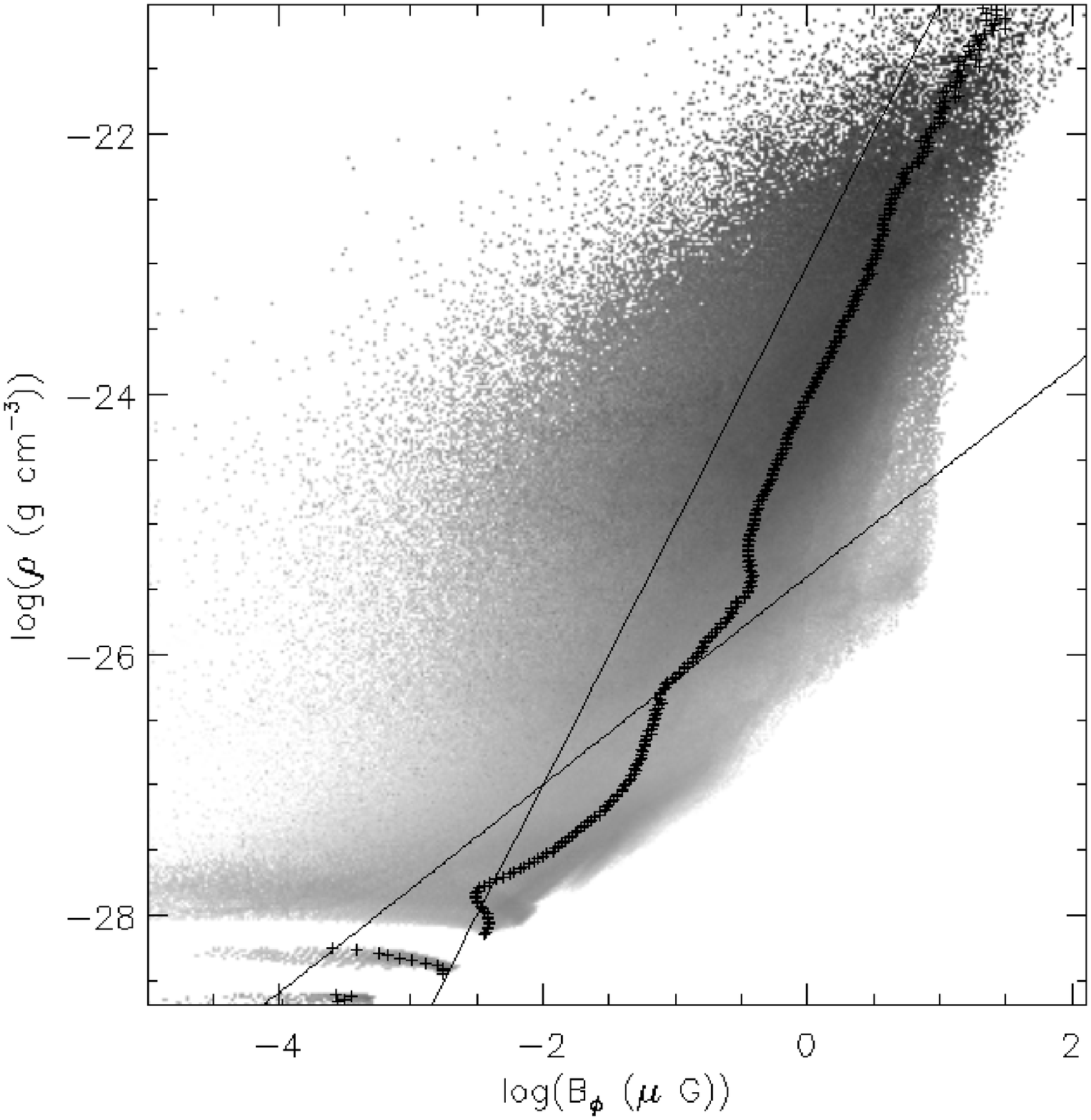}
    \includegraphics[height=.3\textheight]{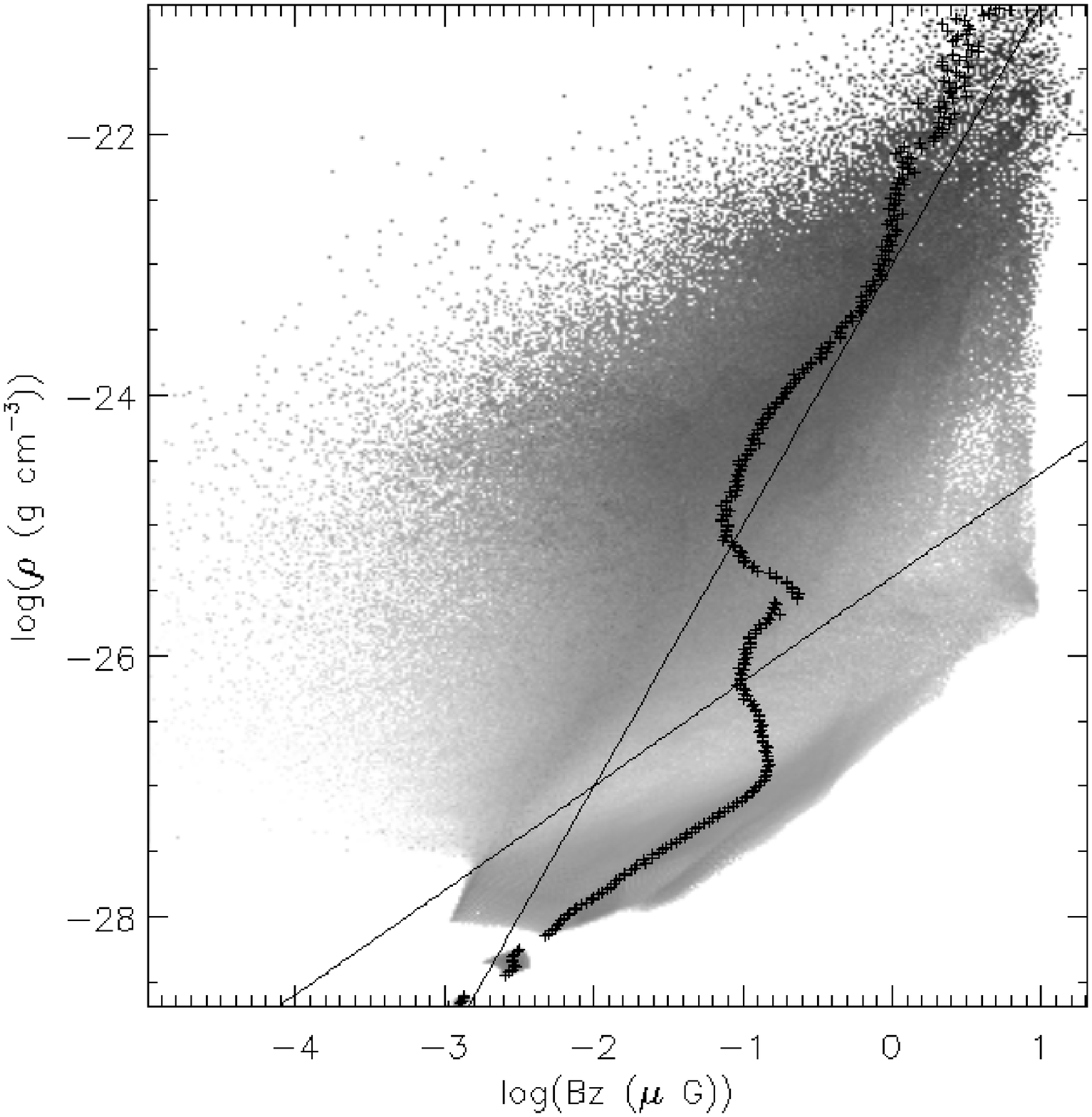}
    \caption{Top: mass-weighted joint density-$B_{\phi}$ PDF at $t=1.088$ Gyr. Bottom: mass-weighted joint density-$B_z$ PDF at $t=1.088$ Gyr. The color scale and binsize are the same as that of Fig. \ref{tb}. The plus sizes are the averaged $B_y(B_z)$ in every density bins. The solid straight lines are $\rho\propto B_\phi^{2}$ and $\rho\propto B_\phi^{0.8}$ in the top panel and $\rho\propto B_z^{2}$ and $\rho\propto B_z^{0.8}$ in the bottom panel.} \label{rhobphi}
\end{figure}

\begin{figure}
 % \plotone{f4.eps}
  \includegraphics[height=.25\textheight]{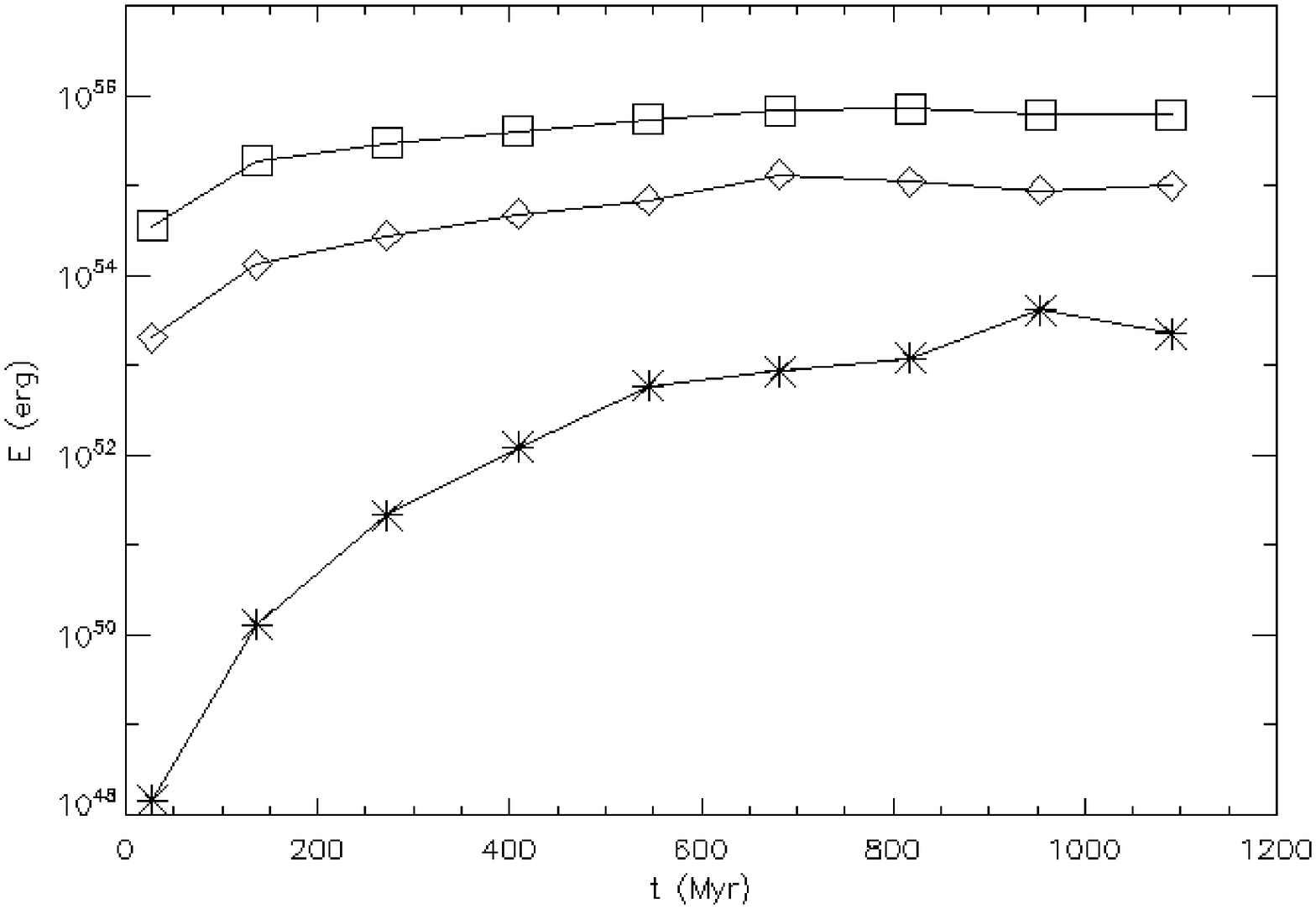}
  \includegraphics[height=.25\textheight]{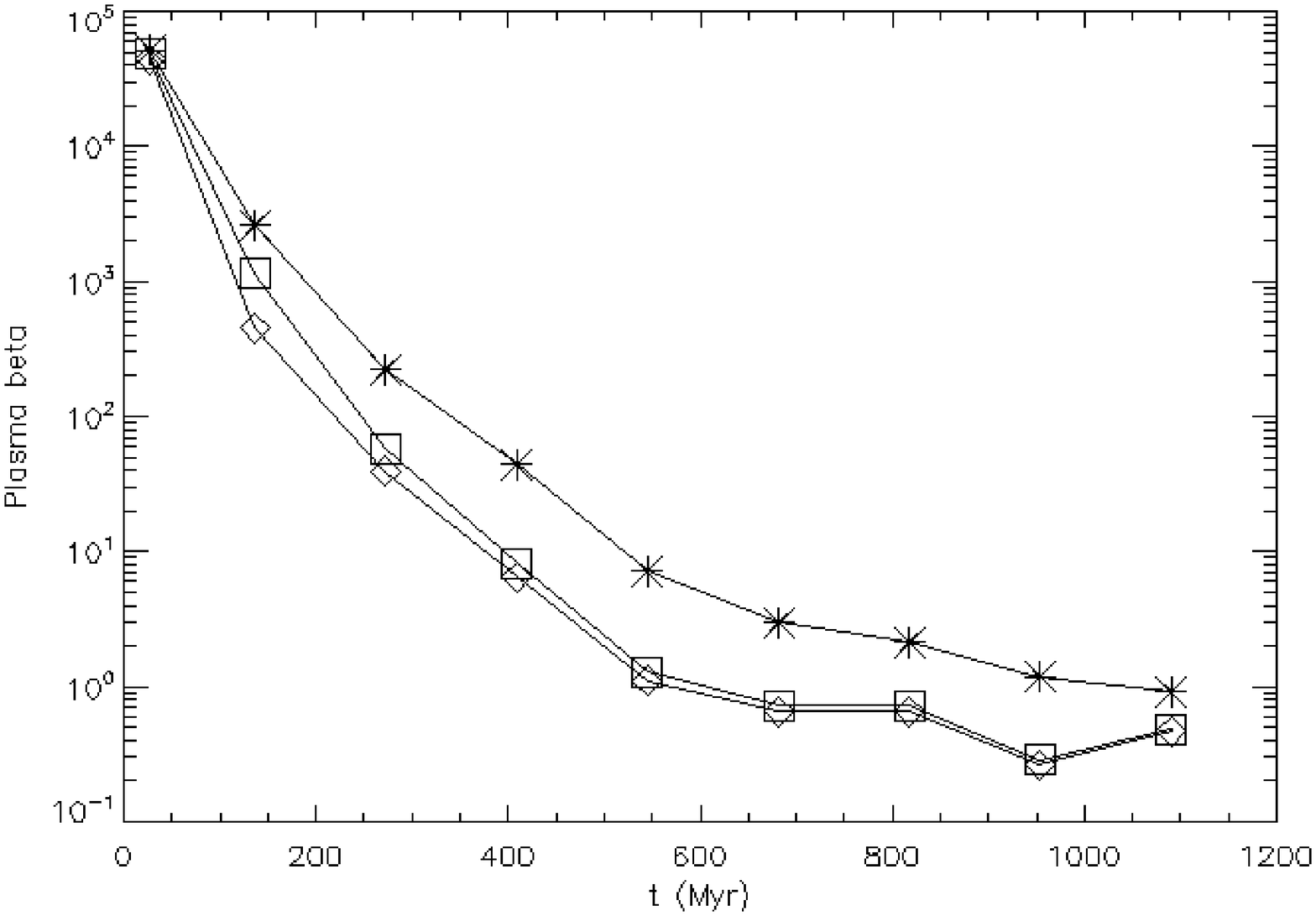}
    \caption{Top: time evolution of total disk gas kinetic energy (squares), thermal energy (diamonds) and magnetic energy (asterisks). Bottom: time evolution of mass-weighted average plasma beta for total gas (squares), cold gas (diamonds) and warm gas (asterisks).} \label{energy}
\end{figure}

\subsection{The Growth and Structure of the Magnetic Field}

Comparing the density projections of the Hydro and MHD cases in
Fig.~\ref{proj}, we can see that the low filaments are more coherent
in the MHD case. Correspondingly, the MHD disk has less low density
regions. By comparing the velocity field to magnetic field in
Fig.~\ref{bfield}, we can see that the large-scale velocity field and
magnetic field are aligned in many places of the disk. This may be a
result of self-organization in MHD turbulence by dynamic alignment of
velocity and magnetic fields \citep{biskamp93}. This implies that
flows along field lines may be important for the formation of clouds
\footnote{We thank Ralph Pudritz for alerting us to this point.}.

The edge-on slice of magnetic field lines shows that at large scale the poloidal field looks like a split monopole. This is the result of an initial uniform magnetic field along the rotation axis being dragged in by the collapse. The physics of magnetic field amplification in MRI, i.e., the stretching of magnetic field lines increases their energy density at the expense of differential rotation, must also work here. From the face-on slices in Fig.~\ref{bfield}, it can be seen that in the horizontal direction the magnetic field amplification is restricted mainly to the disk. But from the edge-on slice, it is clear that the field amplification extends significantly above the disk plane. As a result, there are highly magnetized bubbles with $\beta\sim10^{-2}$ around the galactic disk. Those low beta bubbles form because density decreases sharply above the disk plane but the magnetic field strength decreases much slower. Similar phenomena are well-known in the context of black hole accretion disks (e.g. Stone \& Pringle 2001). The existence of such low beta bubbles implies that the magnetic pressure may significantly influence the dynamics above the disk plane. Thus halo fields built in this way may be crucial for halo-disk matter interchange \citep{mckee77, norman89, gomez92}, in mergers with other galaxies \citep{toomre72, white78}, ram-pressure stripping \citep{gunn72}, galaxy harassment \citep{moore96} and galaxy strangulation \citep{larson80}.

Fig. \ref{tb} shows the mass-weighted joint temperature-B probability distribution function (PDF). From it we can see that the cold gas has a typical magnetic field strength $\sim 1-10$ $\mu$G. This is remarkably consistent with observations in the Milky Way \citep{crutcher99}. Since $\beta\propto B^{-2}$, this implies that in cold gas, $\beta$ fluctuates by two order of magnitudes. Thus the average $\beta$ gives only limited information for the magnetization of cold gas. Similar result has also been found in local MHD simulations of molecular cloud \citep{tilley07}. Furthermore, Fig.~\ref{tb} also shows that magnetic field strength is similar in warm and cold medium (see also Fig.~\ref{bfield}). Then since thermal pressure is different in those two phases, this results in the vast difference of plasma beta in those two phases.

Fig. \ref{rhob} shows the mass-weighted joint density-B PDF. It shows that the density-B relation roughly follows power law relation $B\propto rho^k$. For the high density disk gas, $k=0.5$. This implies that the Alfven speed is similar in those density ranges. This is also consistent with the observations of $B-\rho$ relation in dense molecular gas \citep{crutcher99}. The scatter in both observational data and our result are quite large. For the low density halo gas, $k=1.25$. This results from the rise-up of the low beta bubble. 

To see how the magnetic field components scale with density, Fig.~\ref{rhobphi} shows the mass-weighted joint density-$B_\phi$ and density-$B_z$ plots. It can be seen that both components have similar density-dependence. In the high density gas, azimuthal magnetic component dominates over the vertical component, similar to what was found in shearing box simulations of MRI \citep{hawley96}. In the low density bubbles, vertical magnetic field dominates over the azimuthal component. Interestingly, this is also consistent with radio polarization observations of nearby edge-on galaxies \citep{krause06}.

\begin{figure}
 % \plotone{f4.eps}
  \includegraphics[height=.3\textheight]{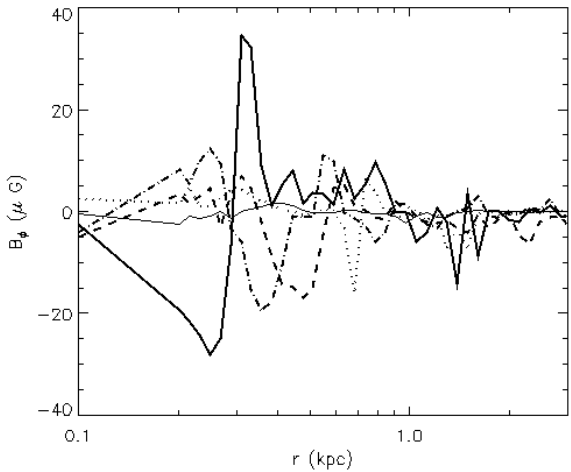}
    \caption{Mass-weighted average of the toroidal magnetic field strength at $t=272$ Myr (thin solid line), 544 Myr (dotted line), 680 Myr (dashed line), 816 Myr (dash-dotted line) and 1.088 Gyr (solid line)} \label{Bz}
\end{figure}

\begin{figure}
 % \plotone{f4.eps}
  \includegraphics[height=.3\textheight]{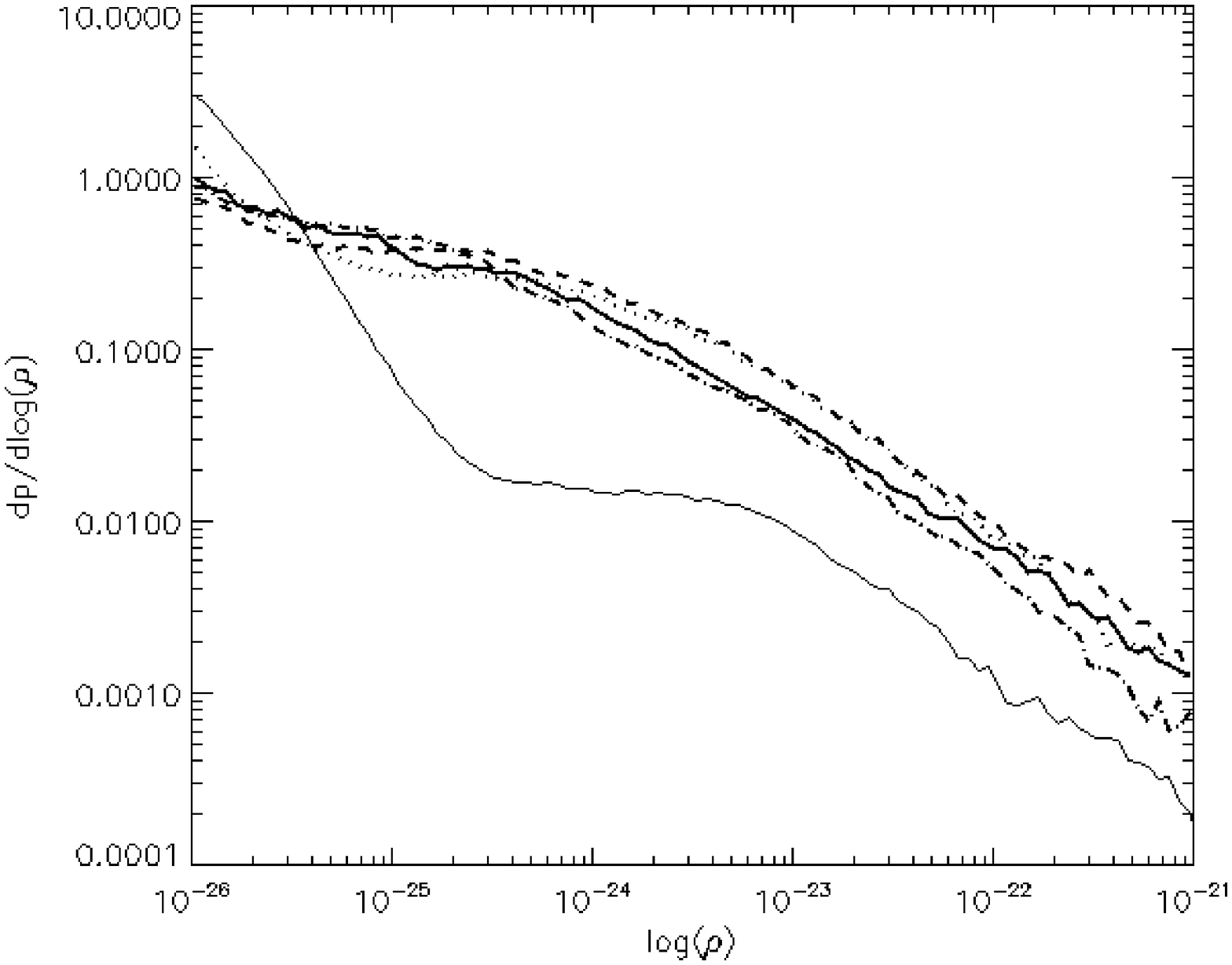}
     \includegraphics[height=.3\textheight]{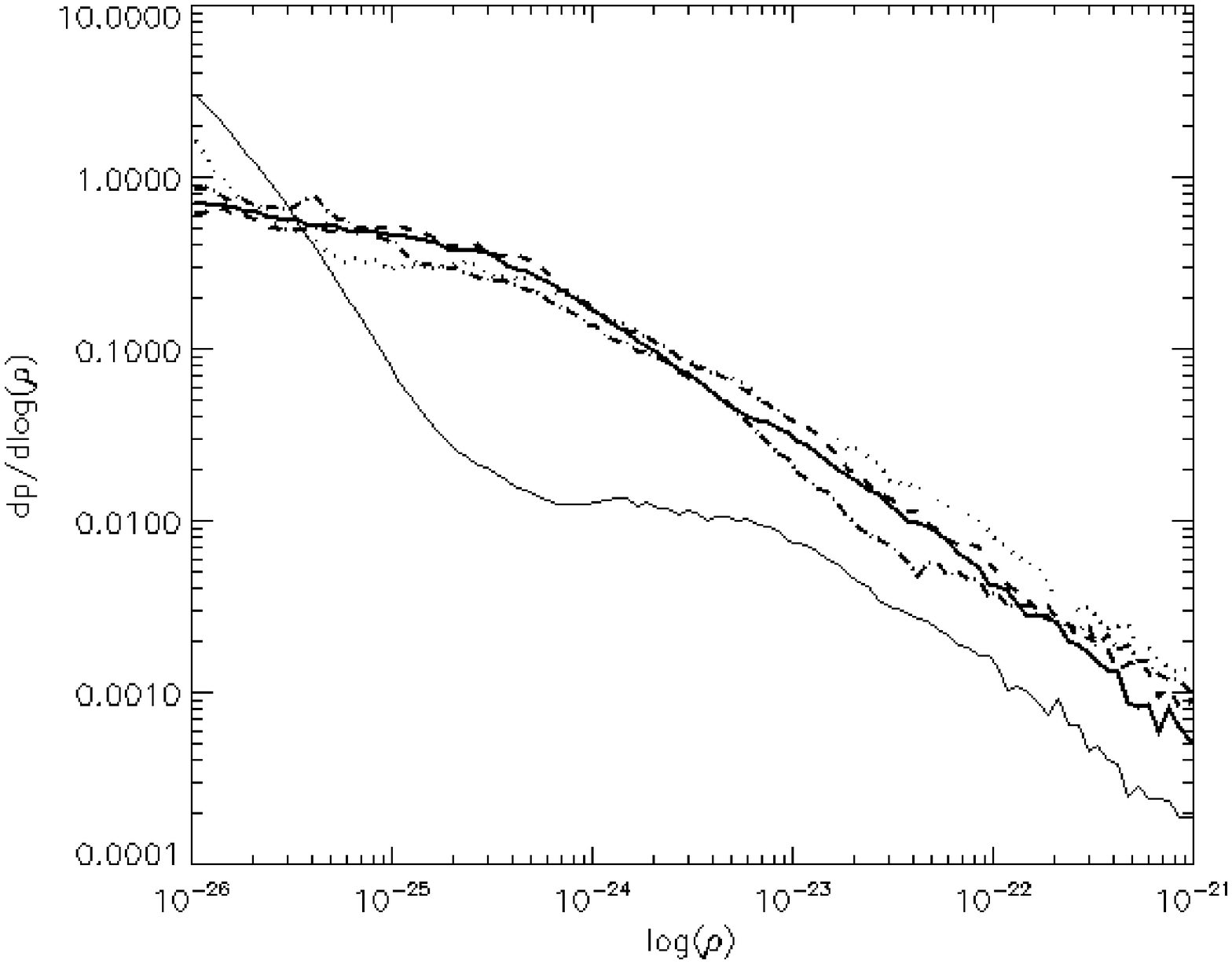}
    \caption{Density PDF for Hydro (top) and MHD (bottom) runs at $t=272$ Myr (thin solid line), 544 Myr (dotted line), 680 Myr (dashed line), 816 Myr (dash-dotted line) and 1.088 Gyr (solid line).} \label{pdf2}
\end{figure}

To examine the amplification history of the magnetic field, the top
panel of Fig.~\ref{energy} shows the time evolution of the total
kinetic energy, thermal energy and magnetic energy for the disk
gas. The magnetic energy increases exponentially by about five orders
of magnitude in the first 600 Myr. Afterwards it remains invariant
with a saturated magnetic field strength in the disk. Some of the
newly amplified field after saturation will rise above the disk,
magnetizing the halo and another fraction of it will be lost by
dissipation. Furthermore, after saturation, the magnetic energy is
still three orders of magnitude smaller than the total kinetic
energy. This explains in part why in our simulation magnetic fields
did not influence significantly the global disk properties. It could
happen that at higher resolution magnetic fields energy can be
amplified to even larger value due to unresolved small-scale dynamo
effects and may influence the disk more significantly. Nevertheless, it is
encouraging that with the current resolution, the magnetic field strength
in the cold gas already is very similar to the observed values in the
Milky Way \citep{crutcher99}.  This is different from the result of dynamo simulations
in MRI-driven turbulence where the total magnetic energy in the
saturated state is generally larger than the kinetic energy
\citep{hawley96}. The primary differences are that turbulence in
galactic disk is mainly driven by self-gravity. Furthermore, the
simulated galactic disk has a multi-phase medium. The cold phase controls the
saturation. And the warm phase still has a subdominant
magnetic field in the saturated state. This also contributes to the
subdominance of magnetic energy.

The bottom panel of Fig.~\ref{energy} shows the time evolution of the
mass-weighted averaged logarithm of  plasma beta for all gas, cold gas
and warm gas in the disk. For $t<600$ Myr, the plasma beta in the cold
gas quickly deceases from $10^5$ to one and then remains around
unity. The magnetic field in the warm phase is always highly
supercritical. This evolution behavior suggests that magnetic field
amplification in disk galaxies is a self-regulated process. After the
plasma beta reaches one in the cold gas, the magnetic fields become
dynamically important and buoyant \citep{parker66}. Further growth of
magnetic field is quenched in the disk. This could be because of
insufficient resolution leading to too large dissipation, yet our $52$
pc resolution run gives very similar result. On the other hand, we can
see clearly that magnetic fields begin to rise above the disk after
saturation, forming the low beta bubbles. This suggests that at least
up to $1$ Gyr, buoyancy is a relevant mechanism of removing magnetic
flux from the disk. To further confirm this picture, we have run
simulations that start with a ten times higher initial magnetic field,
i.e. $10^{-8}$ G. They reach the self-regulating phase earlier and
then drive more low beta bubbles magnetizing the halo in agreement
with the preceding interpretation.

The thick line in Fig.~\ref{Bz} shows the mass-weigthed toroidal
magnetic field at $t=1.088$ Gyr. There are regular patterns of field
reversal at late times. This is a consequence of field amplification
by differentially rotating disks \citep{parker79}. Fig.~\ref{Bz} also
shows the time evolution of the mass-weigthed toroidal magnetic
field. It can be seen that at early times the averaged toroidal fields
are small, before they are amplified by disk rotation. The amplitude
generally decreases with radius because the rotation frequency
decreases. A perhaps more surprising phenomenon occurred between
$t=816$ Myr to $t=1.088$ Gry, when the toroidal field reversed almost exactly symmetrically.

\begin{figure}
 % \plotone{f4.eps}
  \includegraphics[height=.33\textheight]{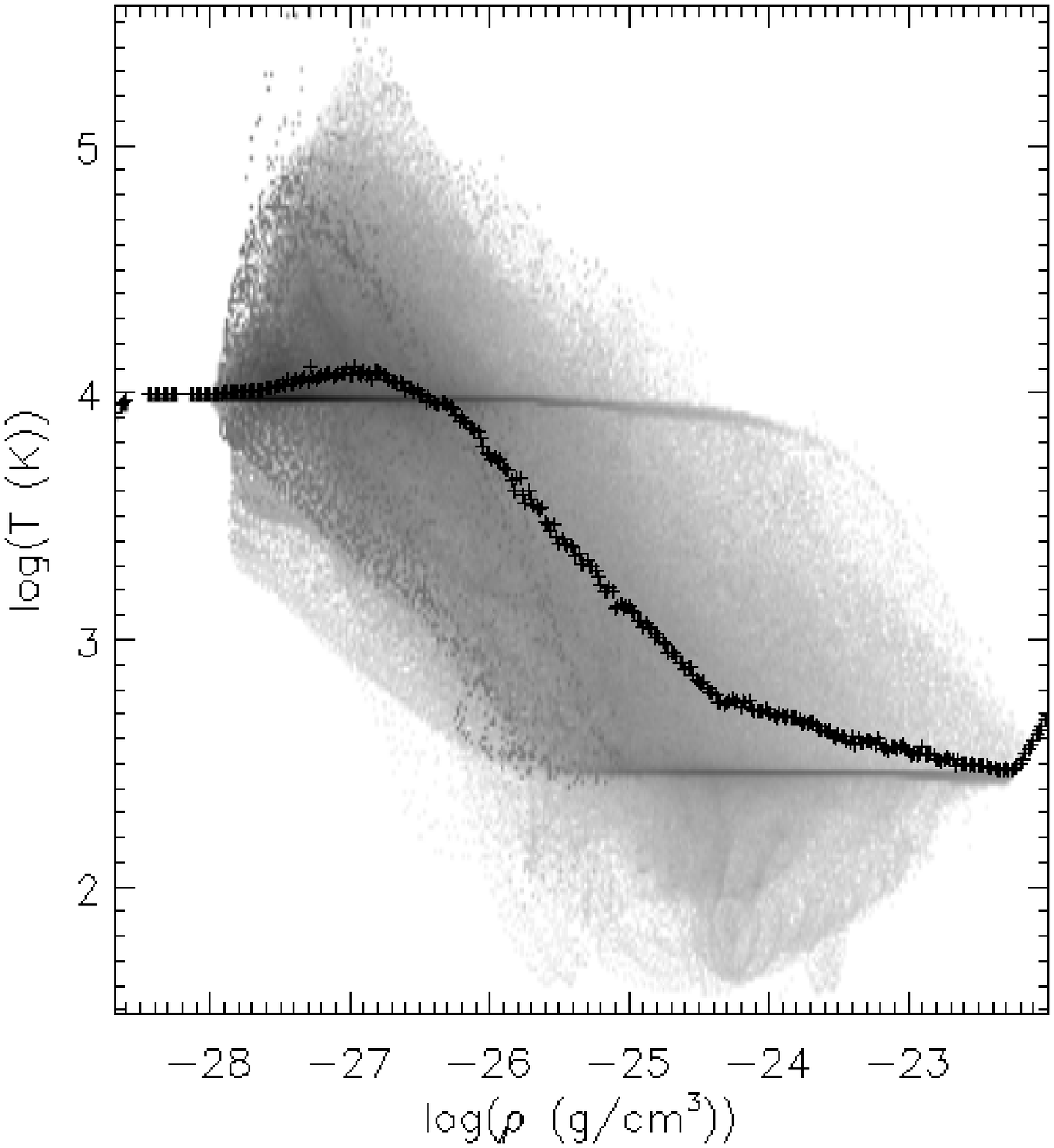}
   \includegraphics[height=.33\textheight]{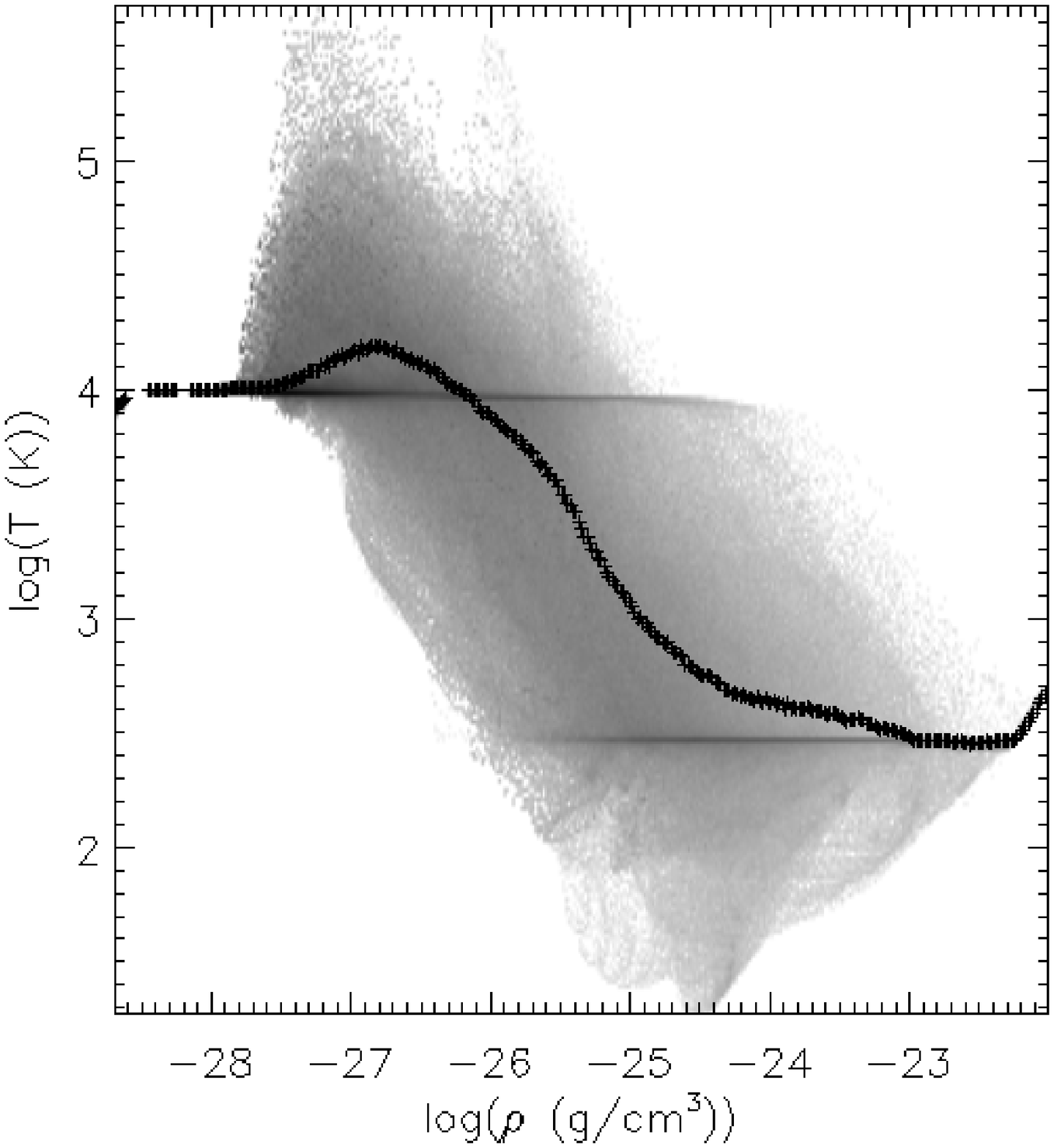}
    \caption{Joint Temperature-Density PDF for Hydro (top) and MHD (bottom) runs at $t=1.088$ Gyr.The color scale and binsize are the same as that of Fig. \ref{tb}. The plus signs are the averaged temperature in every density bins.} \label{trho}
\end{figure}

\begin{figure*}[!]
 % \plotone{f4.eps}
  \includegraphics[height=.35\textheight]{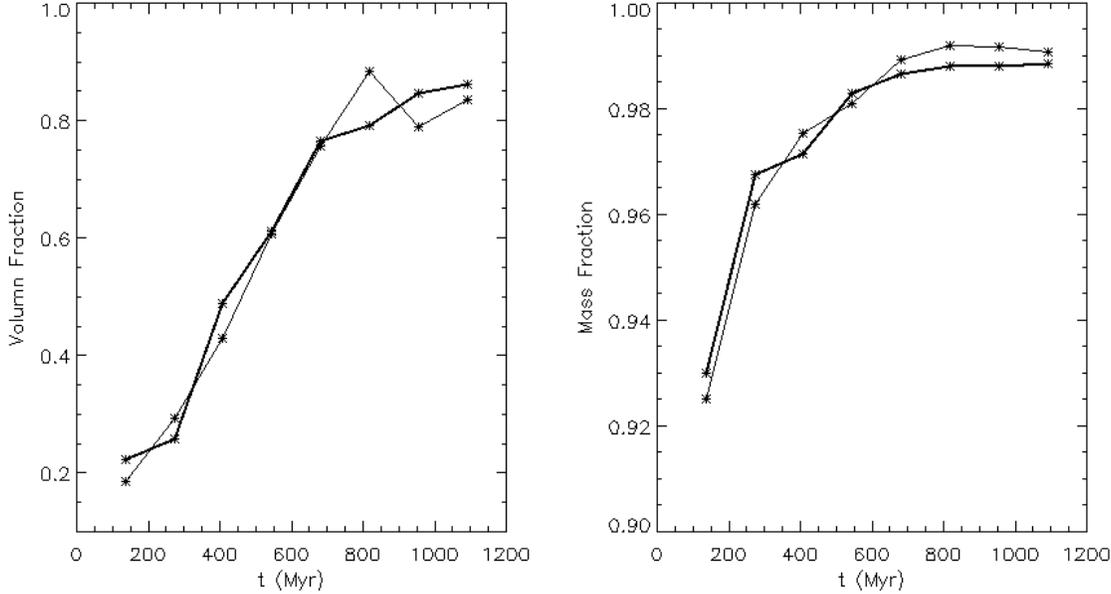}
    \caption{Time evolution of the volume and mass fraction of cold ($T < 10^3$ K) gas for the Hydro (thin line) and MHD (thick line) runs.} \label{frac}
\end{figure*}

%\begin{figure}
 % \plotone{f4.eps}
  %\includegraphics[height=.3\textheight]{pdf.eps}
    %\caption{Density PDF for Hydro (thin) and MHD (thick) run at $t=1.088$ Gyr.} \label{pdf}
%\end{figure}

\subsection{Statistical Properties of the ISM}
\label{ism}

The density PDF is of interest with regards to star formation since star formation is thought to be determined by the high-density tail of the PDF \citep{elmegreen02, mckee05, wada07}. A higher order (though not necessarily small!) effect is the spatial distribution of mass. When calculating the density PDF, we considered only gas in the disk. Fig.~\ref{pdf2} shows the time evolution of this density PDF for both runs.  It can be seen that the PDFs for the two cases are similar. The MHD calculation has slightly less volume in dense gas. In both cases the PDF reaches a quasi-stationary state within less than $\sim500$ Myr. This quick establishment of a quasi-stationary PDF suggests that the gravity-driven turbulence in the disk reached a quasi-stationary and self-regulated state. This phenomenon, if robust even including supernova feedback, may play a crucial role in explaining the the observed universal relation between star formation rate and gas density \citep{kennicutt98, wong02}. Our temperature floor is a feedback on the flow that in reality will be dominated by HII regions and supernova feedback.

Our density PDFs cannot be fitted by log-normal distributions. At
least, they do not show the typical turn-over of a log-normal
distribution. This is different from previous simulations of disk
galaxies which claimed that log-normal is a good fit to the density
PDF \citep{tasker06, tasker07, wada07}. The crucial difference could
be that those studies start from an isolated disk while in our case,
the disk is built inside-out dynamically. Thus anytime in the
evolution of the disk galaxy ($<1$ Gyr) we have simulated, there is
low density gas reaching the disk from outside. It is possible that if
we continue to evolve the disk long enough that after the initial halo
gas all fell down to the disk it may also obtain a log-normal
PDF. However, for a realistic cosmological environment, infall should
never stop completely.

Fig.~\ref{trho} shows the joint temperature-density PDF. Above $\rho >
8\times10^{-23}$~\den, the temperature is determined by the floor
value which was introduced to ensure the calculation resolves the
Jeans length. So the temperature of the gas can at most be trusted
below this density. From this figure, it is also clear that the ISM develops a two-phase structure with a warm $T\sim10^4$ K and a cold $T\sim300$ K phase. However, unlike the originally quasi-stationary picture of a two-phase ISM \citep{field65}, where the phase is characterized by a density and temperature determined by the balance of heating and cooling, in our case there are three order of magnitude density fluctuations in both cold and warm gas. This is due to the turbulent nature of heating in our simulation as discussed above. However, in both runs, one can still identify a good minimum density $\sim 10^{-26}$ \den for the cold phase. 

Fig. \ref{frac} shows the evolution of the volume and mass fractions of cold gas in both runs. It can be seen that the amount of cold gas is monotonically increasing in both volume and mass fractions. After $600$ Myr, both runs reach a quasi-stationary volume fraction of $\sim0.8$ and mass fraction $\sim0.99$.

\begin{figure}
 % \plotone{f4.eps}
  \includegraphics[height=.3\textheight]{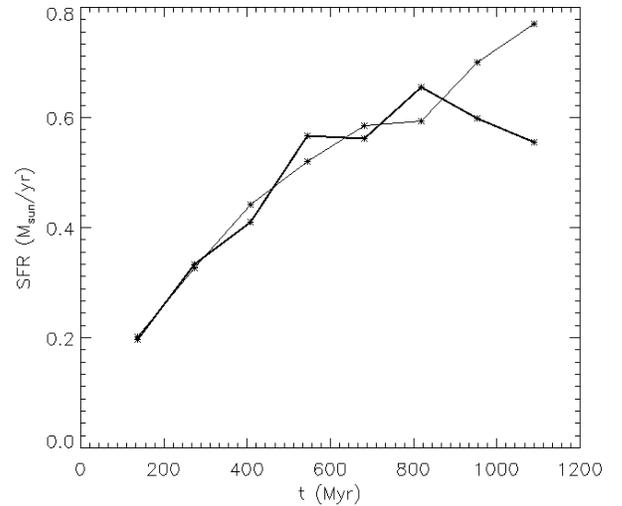}
    \caption{Star formation history of Hydro (thin line) and MHD run (thick line) with $\rho_{cr}=10^{-22}$ \den and $f_s=0.01$.} \label{sfh}
\end{figure}

\begin{figure}
 % \plotone{f4.eps}
  \includegraphics[height=.3\textheight]{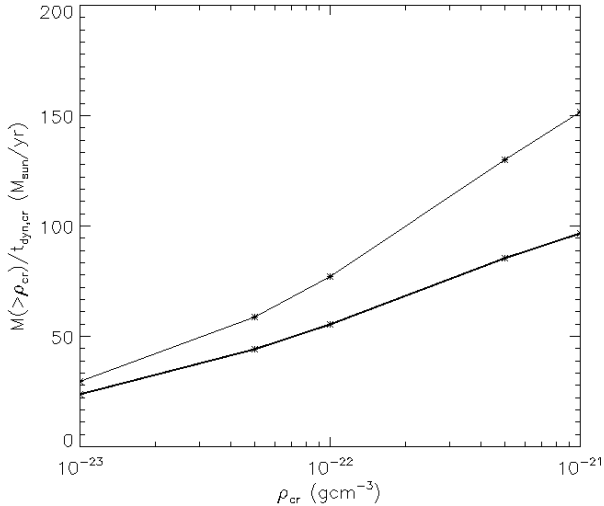}
    \caption{Ratio of $M(\rho>\rho_{cr})/t_{dyn,cr}$ as a function of $\rho_{cr}$ computed from the density PDF for Hydro (thin line) and MHD (solid line) runs at $t=1.088$ Gyr.} \label{mt}
\end{figure}

\subsection{Estimated Star Formation History}

Although we did not model star formation explicitly, from the density PDF, we can estimate the star formation rate (SFR) from \citep{elmegreen02, mckee05, wada07}
\begin{equation}
{\rm SFR} = f_s{M(>\rho_{cr})\over t_{dyn,cr}} \ ,
\end{equation}
where $M(>\rho_{cr})$ is the total gas mass with density $\rho>\rho_{cr}$ which can be computed from the density PDF, $t_{dyn,cr}=1/(G\rho_{cr})^{1/2}$ is the dynamical time at $\rho_{cr}$ and $f_s$ is a constant specifying star formation efficiency (SFE). 

Using $\rho_{cr}=10^{-22}$ \den and $f_s=0.01$, Fig. \ref{sfh} shows
the star formation history (SFH) of the two simulations. It can be
seen that while the SFR in the Hydro run is constantly increasing, in
the MHD case,it reaches a peak at $t\approx800$ Myr and then
decreases. Since $t\approx800$ Myr is shortly after the time when the
average beta in the cold gas reached unity, this suggests that
magnetic forces provide further support in the cold gas after the
amplification saturated. 

%Naively one would expect $f_s$ to depend on $\rho_{cr}$. But recently \citet{tan07} argued that current data suggests a very weak dependence of $f_s$ on density in the range $10-10^5$ cm$^{-3}$. 

Illustrating how sensitive the estimated SFR depends on $\rho_{cr}$,
Fig.~\ref{mt} shows $M(>\rho_{cr})/t_{dyn,cr}$ as a function of
$\rho_{cr}$. It can be seen that in both cases it does depend on
$\rho_{cr}$. Thus to get the same SFR, the SFE $f_s$ should decrease
in those densities ranges. However, it is interesting that the effect
of magnetic field is to make the dependence weaker. From $n=10$
cm$^{-3}$ to $n=10^3$ cm$^{-3}$, in the MHD case
$M(>\rho_{cr})/t_{dyn,cr}$ increases by only a factor of $4$. 

The current analysis of the SFR may be a poor predictor of what happens when molecular cloud and star formation and feedback are included more realistically. However, the analysis in this section does suggest that after the magnetic growth saturates in the disk, magnetic fields can significantly influence GMC formation which occurs in the strongly magnetized cold and dense gas.

%Note that this is not in direct contradiction with the conclusion of \citep{tan07} who based on observational data argued that SFE should be constant over a even larger density range. The reason why SFE decreases with density in our simulation is that a fraction of gas in this density range is not self-gravitating.

%This implies that if the conclusion of \citet{tan07} is correct and our simulation applies to the system they have considered, then additional physics than that included in our simulation should be responsible for establishing the weak dependence of $f_s$ on $\rho_{cr}$. Probably small-scale mechanical and/or radiative feedbacks from massive stars are responsible for this \citep{li06, tom07}.

\section{Discussion}

Using the Spitzer values for viscosity and resistivity, the magnetic Prandtl number of the warm ISM is $\sim10^{11}$. Thus, small-scale dynamos are expected to develop. However, those small-scale dynamo effects are not captured by our resolution. Nevertheless, previous simulations \citep{schekochihin04} have shown that for very high magnetic Prandtl number MHD turbulence, the growth of the small-scale magnetic fields is controlled by the rate of strain at the viscous scale and resulted in magnetic energy tending to pile up at the much smaller resistive scale. This is because for small scale dynamos, the magnetic field seems to retain its folded structure in saturation with direction reversals at the resistive scale \citep{schekochihin02}. So small-scale dynamo processes may not be important for the large-scale growth of galactic magnetic field. At least, the situation is uncertain. Even if small-scale dynamos would contribute some fraction to the large-scale growth of the galactic magnetic field, our simulations of magnetic field growth would be a lower limit. Since we find that the field can be amplified to the observed value quickly, for the phenomenology of galaxy formation, it seems sufficient not resolve small-scale dynamos at the current level of details.

\citet{avillez05} have discussed the results of 3D AMR MHD simulations of a local patch of ISM with sufficient vertical height to include the effect of halo-gas circulation. They find that gas transport to the halo is not significantly influenced by the magnetic field and they do not find the low beta bubble as seen in our simulations. We think the crucial difference of our simulation from theirs is the presence of global differential rotation, which dominates magnetic field amplification in our simulations. 

An important limitation of the presented calculations is that we
neglected supernova feedback while their dynamics could be important
for amplifying the magnetic field in the ISM \citep{ferriere92}. More
importantly, supernova feedback will create the hot phase of the ISM
\citep{mckee77}, which may play a crucial role in regulating star
formation \citep{silk97}. Future work needs to include this supernova
feedback to study the importance of superbubbles on the structure of
ISM, global star formation and magnetic field amplification. Another
limitation is that we neglected cosmic ray pressure, which may be
dynamically important. For example, cosmic ray pressure arising from
the Alfven wave instability \citep{kulsrud69} may prevent the downsliding of matter along the magnetic field line \citep{rafikov00}. 

\section{Conclusions}

The simulations discussed in this work have shown that the amplification of magnetic fields and its large-scale topology is a natural part of the galaxy formation process. Our main findings can be summarized as:

1. In the initial stage of the build-up of a galactic disk, the
magnetic field is very weak and does not significantly affect the dynamics of disk formation. 

2. In a self-gravitating disk, vortex modes driven by the self-gravity
of the gas grow rapidly and trigger the first generation of GMCs formation. 

3. In the cold and warm phase created by turbulent heating, the pressure is not isobaric. Temperature and density fluctuates by about three orders of magnitude in both phases.

4. The dynamical formation of a galactic disk results in a global density PDF that cannot be fitted by a log-normal function.

5. Differentially rotating galactic disks amplify a seed field of $10^{-9}$ G to microgauss levels in $\sim500$ Myr. 

6. The growth of galactic magnetic fields is a self-regulated
process. The saturation is reached first in the cold phase. After
saturation, the field strength agrees with the observations of Milky
Way magnetic fields. The cold phase, the magnetic field strength
fluctuates by about an order of magnitude which gives larger than two
order of magnitude fluctuations in the associated plasma beta. The
saturated total magnetic field energy is three orders of magnitude
smaller than the total kinetic energy in the case presented here.

7. The saturation results in highly magnetized material around galactic disks at all but the earliest cosmic epochs. The halo magnetic fields may significantly influence the halo-gas interactions, accretion of gas in disk galaxies, galaxy mergers and the interactions of galaxies with their environment.

8. After saturation, the toroidal field in the disk dominates over vertical components while in the magnetized halo, vertical components dominate over toroidal components.

9. Large scale magnetic field and velocity field are aligned at many
places of the disk. This implies that cloud formation is likely
channeled by flow along field lines.

10. After saturation, the magnetic field strength is similar in the cold and warm medium.

11. Global toroidal field reversals develops naturally in a differentially rotating disk.

12. Magnetic fields can suppress star formation by providing additional pressure support in the cold gas after saturation.

We have presented the first global numerical experiments
studying magnetic field amplification and evolution during the
formation of disk galaxies. We find it encouraging that many of the
aspects discussed here are consistent with evidence from observations
(e.g. compare with Beck 2007). Being able to include more of the
relevant physics such as molecular cooling, radiation and supernova
feedback a realistic treatment of star formation, and a consistent
treatment of cosmic ray acceleration transport, will render the next
generation of numerical models useful for a more comprehensive
understanding of the physics of galaxies than the current state of the
art. 

\acknowledgments

We thank Zhi-Yun Li and Ralph Pudritz for very useful comments and
suggestions on the draft. We
also benefitted from useful discussions with Chris McKee and Axel
Brandenburg. We are grateful for the hospitality of KITP and our
interactions with the participants of the program ``Star Formation
through Cosmic Time'' when this work was carried out. We were partially
supported by NSF CAREER award AST-0239709 and NSF Grant
No. PHY05-51164. P. W. acknowledges support by an Office of Technology
Licensing Stanford Graduate Fellowship and a KITP Graduate Fellowship.


\begin{thebibliography}{999}

\bibitem[Anderson et al.(2006)]{anderson} 
Anderson, M., Hirschmann, E.~W., Liebling, S.~L., \& Neilsen, D.\ 2006, Classical and 
Quantum Gravity, 23, 6503

\bibitem[Balbus \& Hawley (1991)]{balbus91}
Balbus, S. A., \& Hawley, J. F., 1991, ApJ, 376, 214

\bibitem[Balsara et al.(2001)]{balsara01a} 
Balsara, D., Benjamin, R.~A., \& Cox, D.~P.\ 2001, \apj, 563, 800

\bibitem[Balsara et al.(2004)]{balsara04} 
Balsara, D.~S., Kim, J., Mac Low, M.-M., \& Mathews, G.~J.\ 2004, ApJ, 617, 339

\bibitem[Balsara et al.(2001)]{balsara01b} 
Balsara, D., Ward-Thompson, D., \& Crutcher, R.~M.\ 2001, \mnras, 327, 715

\bibitem[Banerjee \& Pudritz(2007)]{banerjee07} 
Banerjee, R., \& Pudritz, R.~E.\ 2007, ApJ, 660, 479

\bibitem[Beck(2007)]{beck07} 
Beck, R.\ 2007, \aap, 470, 539

\bibitem[Biermann (1950)]{biermann50}
Biermann, L., 1950, Z.Naturforsch, 5a, 65

\bibitem[Biskamp (1993)]{biskamp93}
Biskamp, D., 1993, \emph{Nonlinear Magnetohydrodynamics}, Cambridge University Press

\bibitem[Bryan \& Norman (1997)]{bryan97} 
Bryan, G.~L. \& Norman, M.~L.\ 1997a, arXiv:astro-ph/9710187

%\bibitem[Blackman \& Field (2000)]{balckman00}
%Blackman, E. G., \& Field, G. B., 2000, ApJ, 534, 984

\bibitem[Chandrasekhar (1961)]{chandra61}
Chandrasekhar, S., 1961, Hydrodynamic and Hydromagnetic Stability, Oxford University Press, Oxford

\bibitem[Crutcher(1999)]{crutcher99} 
Crutcher, R.~M.\ 1999, ApJ, 520, 706

\bibitem[de Avillez \& Breitschwerdt(2005)]{avillez05} 
de Avillez, M.~A., \& Breitschwerdt, D.\ 2005, \aap, 436, 585

\bibitem[Dedner et al. (2002)]{dedner}
Dedner, A., Kemm, F., Kršner, D., Munz, C. -D., Schnitzer, T., \& Wesenberg, M., 2002, J. Comput. Phys., 175, 645

\bibitem[Dolag et al.(1999)]{dolag99} 
Dolag, K., Bartelmann, M., \& Lesch, H.\ 1999, \aap, 348, 351

\bibitem[Dutton et al.(2007)]{dutton07} Dutton, A.~A., van den 
Bosch, F.~C., Dekel, A., \& Courteau, S.\ 2007, \apj, 654, 27

\bibitem[Efstathiou(2000)]{efstathiou00} 
Efstathiou, G.\ 2000, \mnras, 317, 697

\bibitem[Elmegreen (2002)]{elmegreen02}
Elmegreen, B. G., 2002, ApJ, 577, 206

\bibitem[Engargiola et al.(2003)]{engargiola03} 
Engargiola, G., Plambeck, R.~L., Rosolowsky, E., \& Blitz, L.\ 2003, ApJS, 149, 343

\bibitem[En{\ss}lin et al.(2007)]{ensslin07} 
En{\ss}lin, T.~A., Pfrommer, C., Springel, V., \& Jubelgas, M.\ 2007, \aap, 473, 41

\bibitem[Ferguson et al. (1998)]{ferguson98}
Ferguson, A., Wyse, R. F. G., Gallagher, J. S., \& Hunter, D. A., 1998, ApJ, 506, L19

\bibitem[Fermi (1949)]{fermi49}
Fermi, E., 1949, Phys. Rev. 75, 1169

\bibitem[Fermi(1954)]{fermi54} 
Fermi, E.,\ 1954, ApJ, 119, 1

\bibitem[Ferri\'ere (1992)]{ferriere92}
Ferri\'ere,  K., 1992, ApJ, 389, 286

\bibitem[Field (1965)]{field65} 
Field, G.~B.\ 1965, ApJ, 142, 531

\bibitem[Gammie (2001)]{gammie01}
Gammie, C. F., 2001, ApJ, 553, 174

\bibitem[Gomez de Castro \& Pudritz(1992)]{gomez92} 
Gomez de Castro, A., \& Pudritz, R.~E.\ 1992, ApJ, 395, 501

\bibitem[Governato et al.(2007)]{governato07} 
Governato, F., Willman, B., Mayer, L., Brooks, A., Stinson, G., Valenzuela, O., Wadsley, 
J., \& Quinn, T.\ 2007, MNRAS, 374, 1479

%\bibitem[Gruzinov \& Diamond (1994)]{gruzinov94}
%Gruzinov, A., \& Diamond, P. H., 1994, Phys. Rev. Lett., 72, 1651

\bibitem[Gunn \& Gott(1972)]{gunn72} 
Gunn, J.~E., \& Gott, J.~R.~I.\ 1972, ApJ, 176, 1

\bibitem[Guo \& White (2007)]{guo07}
Guo, Q., \& White, S. D. M., 2007, astro-ph/0708.1814

\bibitem[Hawley et al. (1996)]{hawley96}
Hawley, J. F., Gammie, C. F., \& Balbus, S. A., 1996, ApJ, 464, 690

\bibitem[Hennebelle \& Inutsuka (2006)]{hennebelle06} 
Hennebelle, P., \& Inutsuka, S.-i.\ 2006, ApJ, 647, 404

%\bibitem[Kamionkowski (2007)]{kamionkowski07}
%Kamionkowski, M., 2007, astro-ph/0706.2986

\bibitem[Kampakoglou \& Silk(2007)]{silk07} 
Kampakoglou, M., \& Silk, J.\ 2007, \mnras, 380, 646

\bibitem[Kaufmann et  al. (2007)]{kaufmann07}
Kaufmann, T., Wheeler, C., \& Bullock, J. S., 2007, astro-ph/0706.0210

\bibitem[Kennicutt (1998)]{kennicutt98}
Kennicutt, R., 1998, ApJ, 498, 541

\bibitem[Krause et al.(2006)]{krause06} 
Krause, M., Wielebinski, R., \& Dumke, M.\ 2006, \aap, 448, 133

\bibitem[Kravtsov(2003)]{kravtsov03} 
Kravtsov, A.~V.\ 2003, ApJL, 590, L1

\bibitem[Krumholz \& McKee (2005)]{mckee05}
Krumholz, M. R., \& McKee, C. F., 2005, ApJ, 630, 250

%\bibitem[Krumholz \& Tan(2007)]{tan07} 
%Krumholz, M.~R., \& Tan, J.~C.\ 2007, ApJ, 654, 304

\bibitem[Kulsrud \& Anderson (1992)]{kulsrud92}
Kulsrud, R. M., \& Anderson, S. W., 1992, ApJ, 396, 606

\bibitem[Kulsrud et al. (1997)]{kulsrud97}
Kulsrud, R. M., Cen, R., Ostriker, J. P., \& Ryu, D., 1997, ApJ, 480, 481

\bibitem[Kulsrud \& Pearce (1969)]{kulsrud69}
Kulsrud, R. M., \& Pearce, W. P., 1969, ApJ, 480, 481

\bibitem[Kulsrud \& Zweibel (2007)]{kulsrud07}
Kulsrud, R. M., \& Zweibel, E. G., 2007, astro-ph/0707.2783

\bibitem[Kurganov \& Tadmor (2000)]{kurganov00}
Kurganov, A. \& Tadmor, E.\ 2000, J. Comput. Phys., 160, 241

\bibitem[Larson(1981)]{larson81} 
Larson, R.~B.\ 1981, MNRAS, 194, 809

\bibitem[Larson et al.(1980)]{larson80} 
Larson, R.~B., Tinsley, B.~M., \& Caldwell, C.~N.\ 1980, ApJ, 237, 692

\bibitem[LeVeque (2002)]{leveque}
LeVeque, R.~J.\ 2002, {\sl Finite Volume Methods for Hyperbolic Problems}, Cambridge University Press

\bibitem[Li et al. (2006)]{li06}
Li, S., Li, H., \& Cen, R., 2006, astro-ph/0611863

\bibitem[Li et al.(2005)]{li05} 
Li, Y., Mac Low, M.-M., \& Klessen, R.~S.\ 2005, ApJ, 626, 823

\bibitem[Li \& Nakamura(2006)]{zhiyun06} 
Li, Z.-Y., \& Nakamura, F.\ 2006, ApJL, 640, L187

\bibitem[Mac Low et al.(2005)]{maclow05} Mac Low, M.-M., 
Balsara, D.~S., Kim, J., \& de Avillez, M.~A.\ 2005, \apj, 626, 864

\bibitem[Mac Low \& Klessen(2004)]{maclow04} 
Mac Low, M.-M., \& Klessen, R.~S.\ 2004, Reviews of Modern Physics, 76, 125

\bibitem[Mamatsashvili \& Chagelishvili (2007)]{mamatsashvili07}
Mamatsashvili, G. R., \& Chagelishvili, G. D., 2007, MNRAS, 381, 809

\bibitem[Martin \& Kennicutt (2001)]{martin01}
Martin, C. L., \& Kennicutt, R. C., 2001, ApJ, 555, 301

\bibitem[Matsumoto (2006)]{matsumoto} 
Matsumoto, T.\ 2006, PASJ in press, arXiv:astro-ph/0609105

\bibitem[Mellon \& Li (2007)]{mellon07}
Mellon, R. R., \& Li, Z.-Y., 2007, astro-ph/0709.0445

\bibitem[McKee \& Ostriker(1977)]{mckee77} 
McKee, C.~F., \& Ostriker, J.~P.\ 1977, ApJ, 218, 148

%\bibitem[McKee \& Ostriker(2007)]{mckee07} 
%McKee, C.~F., \& Ostriker, E.~C.\ 2007, ARAA, 45, 565

\bibitem[Miniati et al.(2001)]{miniati01} 
Miniati, F., Jones, T.~W., Kang, H., \& Ryu, D.\ 2001, ApJ, 562, 233

\bibitem[Mo et al.(1998)]{mo98} 
Mo, H.~J., Mao, S., \& White, S.~D.~M.\ 1998, \mnras, 295, 319

\bibitem[Moore et al.(1996)]{moore96} 
Moore, B., Katz, N., Lake, G., Dressler, A., \& Oemler, A.\ 1996, Nature, 379, 613

\bibitem[Mouschovias(1987)]{mouschovias87} Mouschovias, T.~C.\ 1987, 
NATO ASIC Proc.~210: Physical Processes in Interstellar Clouds, 453

\bibitem[Navarro et al. (1996)]{nfw}
Navarro, J. F., Frenk, C. S., \& White, S. D. M., 1996, ApJ, 462, 563

\bibitem[Norman \& Ikeuchi(1989)]{norman89} 
Norman, C.~A., \& Ikeuchi, S.\ 1989, \apj, 345, 372

\bibitem[O'Shea et al.(2004)]{o'shea04} 
O'Shea, B.~W., Bryan, G., Bordner, J., Norman, M.~L., Abel, T., Harkness, R., \& Kritsuk, A.\
2004, In "Adaptive Mesh Refinement - Theory and Applications", Eds. T. Plewa, T. Linde \& V. G. Weirs, Springer Lecture Notes in Computational Science and Engineering, 2004

\bibitem[Okamoto et al.(2005)]{okamoto05} 
Okamoto, T., Eke, V.~R., Frenk, C.~S., \& Jenkins, A.\ 2005, MNRAS, 363, 1299

\bibitem[Oren \& Wolfe (1995)]{oren95}
Oren, A. L., \& Wolfe, A. M., 1995, ApJ, 445, 624

\bibitem[Parker (1966)]{parker66}
Parker, E. N., 1966, ApJ, 145, 811

\bibitem[Parker (1971)]{parker71}
Parker, E. N., 1971, ApJ, 163, 255

\bibitem[Parker (1979)]{parker79}
Parker, E. N., 1979, \emph{Cosmical Magnetic Fields: Their Origin and Their Activity}, Oxford University Press

\bibitem[Piontek \& Ostriker (2007)]{piontek07} 
Piontek, R.~A., \& Ostriker, E.~C.\ 2007, ApJ, 663, 183

\bibitem[Pudritz \& Silk (1989)]{pudritz89}
Pudritz, R. E., \& Silk, J., 1989, ApJ, 342, 650

%\bibitem[Robertson et al. (2006)]{robertson06} 
%Robertson, B., Bullock, J.~S., Cox, T.~J., Di Matteo, T., Hernquist, L., Springel, V., \& 
%Yoshida, N.\ 2006, ApJ, 645, 986

\bibitem[Quirk (1994)]{quirk}
Quirk, J.\ 1994, Int. J. Numer. Methods Fluids, 18, 555

\bibitem[Rafikov \& Kulsrud (2000)]{rafikov00}
Rafikov, R. R., \& Kulsrud, R. M., 2000, MNRAS, 314, 839

\bibitem[Rees (2006)]{rees06}
Rees, M. J., 2006, Astron. Nachr., 327, 395

\bibitem[Robertson \& Kravtsov (2007)]{robertson07}
Robertson, B., \& Kravtsov, A., 2007, astro-ph/0710.2102

\bibitem[Rosen \& Bregman (1995)]{rosen}
Rosen, A., \& Bregman, J. N., 1995, ApJ, 440, 634

\bibitem[Sarazin \& White (1987)]{sarazin87}
Sarazin, C. L., \& White, R. E., 1987, ApJ, 320, 32

\bibitem[Schekochihin et al. (2002)]{schekochihin02}
Schekochihin, A. A., Cowley, C. S., Hammett, G. W., Maron, J. L., \& McWilliams, J.C., 2002, New J. Phys., 4, 84

\bibitem[Schekochihin et al. (2004)]{schekochihin04}
Schekochihin, A. A., Cowley, C. S., Taylor, S. F., Maron, J. L., McWilliams, J. C., 2004, ApJ, 612, 276

\bibitem[Shu \& Osher (1988)]{shu88}
Shu, C.-W. \& Osher, S., 1988, J. Comput. Phys., 77, 439

\bibitem[Silk (1997)]{silk97}
Silk, J., 1997, ApJ, 458, 703

\bibitem[Stringer \& Benson(2007)]{stringer07} 
Stringer, M.~J., \& Benson, A.~J.\ 2007, ArXiv Astrophysics e-prints, arXiv:astro-ph/0703380



\bibitem[Spergel et al. (2007)]{spergel07}
Spergel, D. N., et al., 2007, ApJS, 170, 377

\bibitem[Springel et al. (2005)]{springel05} 
Springel, V., et al.\ 2005, Nature 435, 629

\bibitem[Springel \& Hernquist(2005)]{hernquist05} 
Springel, V., \& Hernquist, L.\ 2005, ApJL, 622, L9

%\bibitem[Springel et al. (2006)]{springel06}
%Springel, V., Frenk, C. S., \& White, S. D. M., 2006, Nature, 440, 1137

\bibitem[Springel \& White (1999)]{springel99}
Springel, V., \& White, S. D. M., 1999, MNRAS, 307, 162

\bibitem[Steenbeck et al. (1966)]{steenbeck66}
Steenbeck, M., Krause, F., \& R\"adler, Ke-H, 1966, Z. Naturforsch, 21a, 369

\bibitem[Stone \& Pringle (2001)]{stone01}
Stone, J. M., \& Pringle, J. E., 2001, MNRAS, 322, 461

\bibitem[Strong et al.(2007)]{strong07}
Strong, A. W., Moskalenko, I. V., \& Ptuskin, V. S., 2007, Annu. Rev. Nucl. Part. Sci., 57, 285

\bibitem[Tasker \& Bryan (2006)]{tasker06}
Tasker, E. J., \& Bryan, G. L., 2006, ApJ, 641, 878

\bibitem[Tasker \& Bryan (2007)]{tasker07}
Tasker, E. J., \& Bryan, G. L., 2007, astro-ph/0709.1972

\bibitem[Tassis et al.(2006)]{tassis06} 
Tassis, K., Kravtsov, A.~V., \& Gnedin, N.~Y.\ 2006, astro-ph/0609763

%\bibitem[Tegmark et al. (2006)]{tegmark06}
%Tegmark, M., et al., 2006, Phys.Rev. D74, 123507

\bibitem[Thoul \& Weinberg(1996)]{thoul96} 
Thoul, A.~A., \& Weinberg, D.~H.\ 1996, ApJ, 465, 608

\bibitem[Tilley \& Pudritz(2007)]{tilley07} 
Tilley, D.~A., \& Pudritz, R.~E.\ 2007, MNRAS, 930

\bibitem[Truelove et al.(1997)]{truelove97} 
Truelove, J.~K., Klein, R.~I., McKee, C.~F., Holliman, J.~H., II, Howell, L.~H., \& 
Greenough, J.~A.\ 1997, ApJL, 489, L179

\bibitem[Toomre (1964)]{toomre}
Toomre, A., 1964, ApJ, 139, 1217

\bibitem[Toomre \& Toomre(1972)]{toomre72} 
Toomre, A., \& Toomre, J.\ 1972, \apj, 178, 623

\bibitem[Van Leer (1979)]{van leer79}
Van Leer, B.\ 1979, J. Comput. Phys., 32, 101

\bibitem[Vainshtein \& Ruzmaikin (1971)]{vainshtein71}
Vainshtein, S. I., \& Ruzmaikin, A. A., 1971, Sov. Astron. 15, 74

\bibitem[Wada et al. (2002)]{wada02}
Wada, K., Meurer, G., \& Norman, C. A., 2002, ApJ, 577, 197

\bibitem[Wada \& Norman (2007)]{wada07}
Wada, K., \& Norman, C. A., 2007, ApJ, 660, 276

\bibitem[Wang et al. (2007)]{wang07}
Wang, P., Abel, T., \& Zhang, W., 2007, astro-ph/0703742

\bibitem[White(1978)]{white78} 
White, S.~D.~M.\ 1978, MNRAS, 184, 185

\bibitem[Wong \& Blitz (2002)]{wong02}
Wong, T., \& Blitz, L., 2002, ApJ, 569, 157

\bibitem[Zweibel (2003)]{zweibel03}
Zweibel, E. G., 2003, ApJ, 587, 625

\end{thebibliography}
\end{document}